\documentclass[journal,10pt]{IEEEtran}
\ifCLASSINFOpdf
  \usepackage[pdftex]{graphicx}
  \usepackage{amsthm,amsmath,amssymb}
\usepackage{makecell}
\usepackage{mathrsfs}
  \usepackage{float}
  \usepackage{algorithm}
\usepackage{algpseudocode}
\usepackage{amsmath}
\usepackage{bm}
\usepackage{color}
\usepackage{comment}
\usepackage{hyperref}

\else
\fi
\usepackage{leftidx}
\usepackage{epstopdf}
\usepackage{amsmath}
\usepackage{subfigure}
\usepackage{makecell}
\usepackage{multirow}
\usepackage{enumerate}
\begin{document}

\title{A Multi-modal Intelligent Channel Model for 6G Multi-UAV-to-Multi-Vehicle Communications}

\author{Lu~Bai,~\IEEEmembership{Member,~IEEE}, Mengyuan Lu, Ziwei~Huang,~\IEEEmembership{Member,~IEEE}, and Xiang~Cheng,~\IEEEmembership{Fellow,~IEEE}

	\thanks{The authors would like to thank Xuanyu Liu and Xu Wang for their help in the collection of LiDAR point clouds in AirSim simulation platform.}

\thanks{L.~Bai is with the Joint SDU-NTU Centre for Artificial Intelligence Research (C-FAIR), Shandong University, Jinan, 250101, P. R. China (e-mail: lubai@sdu.edu.cn).}
\thanks{M.~Lu is with the Joint SDU-NTU Centre for Artificial Intelligence Research (C-FAIR), Shandong University, Jinan, 250101, P. R. China, and also with the School of Software, Shandong University, Jinan 250101, P. R. China (e-mail: mengyuanlu@mail.sdu.edu.cn).}
\thanks{Z.~Huang and X.~Cheng are with the State Key Laboratory of Advanced Optical Communication Systems and Networks, School of Electronics, Peking University, Beijing, 100871, P. R. China (email: ziweihuang@pku.edu.cn, xiangcheng@pku.edu.cn).}
}

\markboth{}
{Zeng \MakeLowercase{\textit{et al.}}: Bare Demo of IEEEtran.cls for IEEE Journals}

\maketitle

\markboth{IEEE Transactions on Wireless Communications, vol. xx, no. xx, XX 2025}
{Submitted paper}
		\maketitle
\begin{abstract}
In this paper, a novel multi-modal intelligent channel model for sixth-generation (6G) multiple-unmanned aerial vehicle (multi-UAV)-to-multi-vehicle communications is proposed. To thoroughly explore the mapping relationship between the physical environment and the electromagnetic space in the complex multi-UAV-to-multi-vehicle scenario, two new parameters, i.e., terrestrial traffic density (TTD) and aerial traffic density (ATD), are developed and a new sensing-communication intelligent integrated dataset is constructed in suburban scenario under different TTD and ATD conditions. With the aid of sensing data, i.e., light detection and ranging (LiDAR) point clouds, the parameters of static scatterers, terrestrial dynamic scatterers, and aerial dynamic scatterers in the electromagnetic space, e.g., number, distance, angle, and power, are quantified under different TTD and ATD conditions in the physical environment. In the proposed model, the channel non-stationarity and consistency on the time and space domains and the channel non-stationarity on the frequency domain are simultaneously mimicked. The channel statistical properties, such as time-space-frequency correlation function (TSF-CF), time stationary interval (TSI), and Doppler power spectral density (DPSD), are derived and simulated. Simulation results match ray-tracing (RT) results well, which verifies the accuracy of the proposed multi-UAV-to-multi-vehicle channel model.
\end{abstract}

\IEEEpeerreviewmaketitle

\section{Introduction}
With the wide popularization of the low-altitude economy, intelligent low-altitude transportation has received considerable attention. 
As an emerging sixth-generation (6G) intelligent networked scenario, intelligent low-altitude transportation involves a variety of low-altitude unmanned aerial vehicles (UAVs), all types of vehicles, roadside units, and pedestrians. Considering the security of UAVs and autonomous vehicles as well as more convenient and more efficient information service, more reliable and lower latency communication requirements of 6G intelligent networked low-altitude transportation communications can no longer be efficiently addressed by the conventional communication networks. To better design and analyze the 6G intelligent networked low-altitude transportation communication system, the research on the underlying propagation characteristics and corresponding channel modeling is essential \cite{6ggg}. 

So far, many researchers have worked on UAV-to-ground channel measurement campaigns, channel characteristic analysis, and channel modeling, which can to some extent investigate 6G intelligent networked low-altitude channels. 
The UAV-to-ground channel characteristics at 5.8~GHz in suburban scenarios were measured and analyzed in \cite{mea-1}. 
The UAV-to-ground channel measurement campaigns at 968~MHz and 5~GHz in urban, suburban, and highland scenarios were carried out in \cite{mea-2}, where the single-input multiple-output (SIMO) channel characteristics were investigated, including path loss, Ricean factor, delay spread, and small-scale fading parameters. The authors in \cite{mea-3} conducted the 
UAV-to-ground channel measurement campaigns at 1 GHz and 4 GHz, and analyzed the time non-stationarity in UAV-to-ground channels. 
The spatial channel characterizations of UAV-to-ground channel at 1.8~GHz and 2.5~GHz were respectively investigated based on the measurement campaigns in \cite{mea-4, mea-5}. 
Based upon these channel measurement campaigns and characteristic analysis, extensive UAV-to-ground channel models were proposed. 
According to the electromagnetic wave theory and ray-tracing (RT) technology, the deterministic UAV-to-ground channel models \cite{det1, det2} were proposed. However, the deterministic channel models are limited to certain physical environments. To apply to more diverse UAV-to-ground physical environments, the geometry-based stochastic models (GBSMs), whose parameters can be adjusted with the UAV-to-ground physical environment, are proposed, including regular-shaped GBSMs (RS-GBSMs) and irregular-shaped GBSMs (IS-GBSMs). In UAV-to-ground RS-GBSMs \cite{rs1}, \cite{rs4} the scattering clusters were modeled on two-dimensional (2D) rings, 2D ellipses, three-dimensional (3D) cylinders, and 3D ellipsoids to calculate propagation paths and channel parameters. However, the scattering clusters in RS-GBSMs are too restricted on regular shapes to mimic the high-dynamic UAV channels. Therefore, more suitable and flexible IS-GBSMs \cite{is1}--\cite{is4} were proposed for UAV-to-ground channels.

Nevertheless, the aforementioned UAV-to-ground channel models are limited and insufficient to describe the 6G intelligent networked low-altitude transportation channel. The high-mobility of UAVs and vehicles, the infinity of intelligent agents equipped with communication equipment, and the intricacy of pervasive connectivity bring new challenges for channel modeling. Furthermore, more reliable and lower latency communication requirements of 6G intelligent networked low-altitude transportation communications rely on a more in-depth understanding of the propagation environment and more accurate channel modeling, where the conventional channel modeling method can no longer satisfy. 
Fortunately, in low-altitude intelligent transportation, the multiple intelligent networked UAVs, autonomous vehicles, and roadside units are simultaneously deployed with communication devices and multi-modal sensors. In this case, the communication capability and sensing capability coexist symbiotically, which brings more opportunities for 6G intelligent networked low-altitude transportation channel modeling. 
Inspired by human synesthesia, \emph{Synesthesia of Machines (SoM)} was proposed in \cite{SOM} for the technology development of intelligent multi-modal sensing-communication integration. 
Unlike integrated sensing and communications (ISAC) \cite{ISAC_my}, which focuses on radio-frequency (RF) radar sensing and communications, SoM refers to the intelligent integration of multi-modal sensing and communications, including 
 RF communications,  RF sensing, i.e., millimeter wave (mmWave) radar, and non-RF sensing, i.e., light detection and ranging (LiDAR) and RGB-Depth cameras etc. 
Similar to the way humans sense the environment via multiple organs, i.e., the environmental information obtained by multiple organs is mutually facilitated via biological neural networks, multi-modal sensors and communication devices, i.e., machine sense, can assist mutually and capture more detailed and more accurate environmental information based on machine learning. 
Since channel modeling essentially describes the electromagnetic environment that is closely related to the physical environment, channel modeling with the help of intelligent multi-modal sensing-communication integration has the potential to handle the high-mobility and intricacy of pervasive connectivity in 6G intelligent networked low-altitude transportation communications.

In this paper, inspired by SoM, we explore the mutual facilitation of multi-modal sensing-communication integration in channel modeling of 6G intelligent networked low-altitude transportation communications, investigate the mapping relationship between the physical environment and the electromagnetic space with channel information and LiDAR point clouds, and propose a novel multi-modal intelligent channel model for 6G multiple-UAV (multi-UAV)-to-multi-vehicle communications.
The main contributions and novelties of this paper are summarized below.
\begin{enumerate}
\item To more accurately mimic intelligent networked low-altitude transportation channels, a novel multi-modal intelligent channel model is proposed for 6G multi-UAV-to-multi-vehicle communications. In the proposed model, the impact of terrestrial traffic density (TTD) and aerial traffic density (ATD) is considered for the first time in UAV-to-ground channel modeling. Furthermore, a novel
LiDAR-aided temporal and spatial non-stationarity and
consistent algorithm is developed to simultaneously depict the channel non-stationarity and consistency on the time and space domains and the channel non-stationarity on the frequency domain.

\item To thoroughly explore the mapping relationship between the physical environment and the electromagnetic space in the complex multi-UAV-to-multi-vehicle scenario, a new multi-UAV-to-multi-vehicle cooperative sensing-communication integration (MUMV-CSCI) dataset in suburban forking road scenarios is constructed, including the channel information and LiDAR point clouds. In the constructed dataset, the diversity in the electromagnetic space, i.e., the channels among multiple UAVs and vehicles, and the variety in the physical environment, i.e., the environment under different TTD and ATD conditions, are considered.

\item With the help of sensing information in the physical environment, i.e., LiDAR point clouds, scatterers of multi-UAV-to-multi-vehicle channels in the electromagnetic space can be for the first time divided into static scatterers, terrestrial dynamic scatterers, and aerial dynamic scatterers. In this case, a novel multi-UAV-to-multi-vehicle channel parameter table, e.g., number, distance, angle, and power of dynamic and static scatterers, is developed under different TTD and ATD conditions in the suburban scenario. 

\item The multi-UAV-multi-vehicle channel statistical properties, including time-space-frequency correlation function (TSF-CF), time stationary interval (TSI), and
Doppler power spectral density (DPSD), are derived and simulated. Based on the simulation result, the impact of different TTD and ATD conditions on channel statistics is investigated. Simulation results have close agreement with RT-based results, which verify the proposed multi-UAV-multi-vehicle channel model.
\end{enumerate}

The remainder of this paper is organized as follows. Section~II describes the MUMV-CSCI dataset in the suburban forking road scenario and presents the quantified channel parameters under different TTD and ATD conditions.
In Section~III, a novel multi-modal intelligent channel model for 6G multiple-UAV (multi-UAV)-to-multi-vehicle communications is proposed.
The multi-UAV-to-multi-vehicle channel statistical properties are given in Section~IV. 
Section~V presents the corresponding simulation result, which is further compared with the RT-based result. 
At last, the conclusions are obtained in Section~VI.

\begin{figure*}[!t]
		\centering	\includegraphics[width=\textwidth]{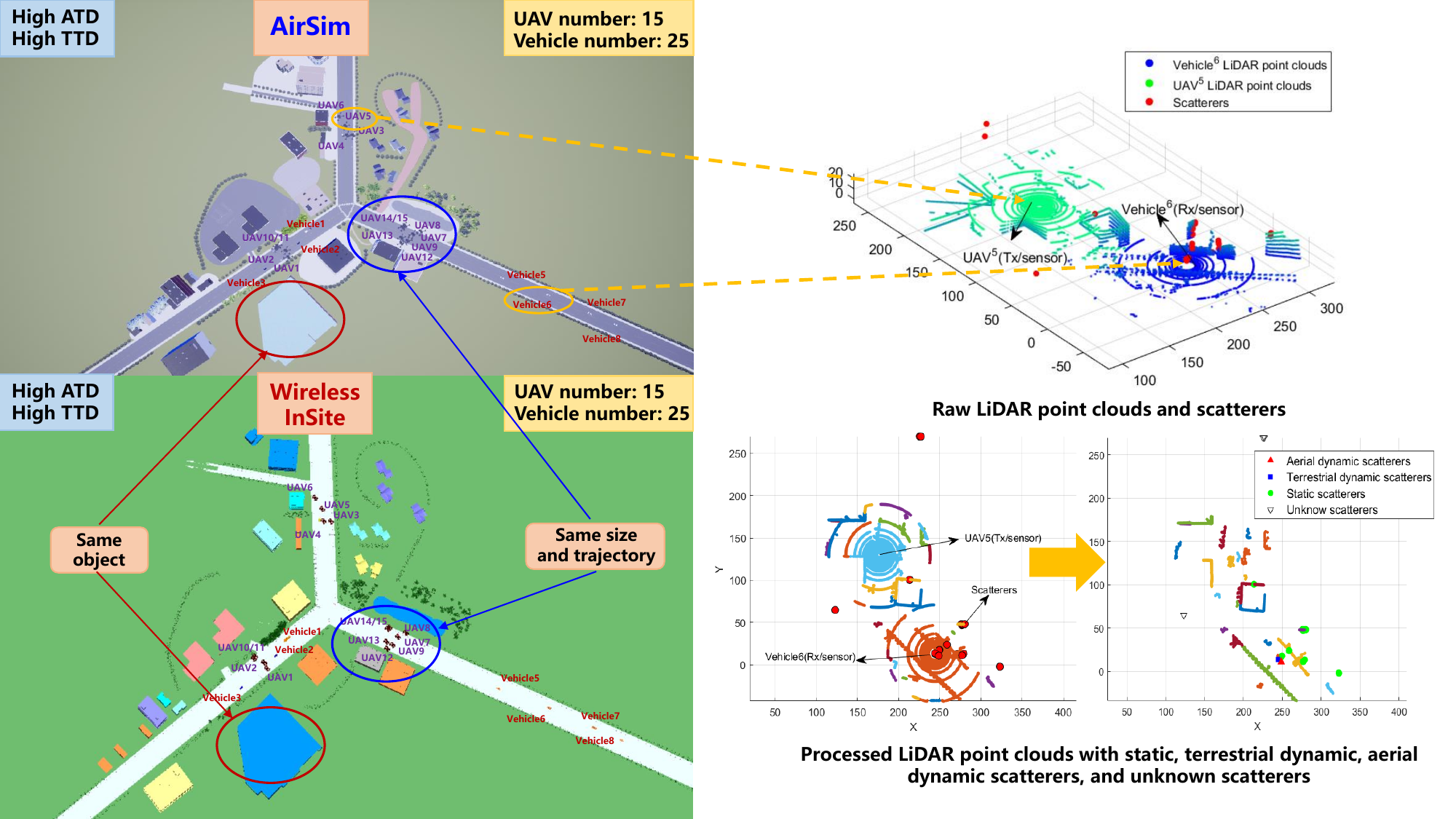}
	\caption{LiDAR point clouds and scatterers in multi-UAV-to-multi-vehicle suburban forking road scenarios under high TTD and ATD conditions in Airsim and Wireless InSite.}
	\label{WI_AirSim}
	\end{figure*}

\section{MUMV-CSCI Dataset and Channel Parameterization  in Suburban Scenario}
The multi-UAV-to-multi-vehicle sensing-communication intelligent integrated measurement campaign is conducted in a suburban forking road scenario. To investigate the impact of traffic density conditions in both terrestrial and aerial areas, the TTD and ATD are developed and the measurement campaign is carried out under different TTD and ATD conditions. 

Since the multi-UAV-to-multi-vehicle channel is highly dynamic, complicated, and changeable, it is significant to explore the static, terrestrial dynamic, and aerial dynamic scatterers. Moreover, investigating the impact of TTD and ATD conditions is essential for the design of 6G multi-UAV-to-multi-vehicle sensing-communication intelligent integrated communication systems.
Nevertheless, the conventional channel measurement campaigns that solely process Doppler information in channels cannot distinguish static, terrestrial dynamic, and aerial dynamic scatterers 
\cite{Lu2024}. To fill this gap, the statistical distributions of key channel parameters related to static, terrestrial dynamic, aerial dynamic scatterers in the multi-UAV-to-multi-vehicle channels are for the first time investigated under high, medium, and low TTD and ATD conditions, which are presented in Table~\ref{Parameter_1}. 
\begin{table}[!t]
    \centering
    \caption{Numbers of LiDAR Point Clouds and Links with Scatterers Under Low, Medium, and High TTD and ATD Conditions}
    \label{WI_link}
    \begin{tabular}{lcc}
        \hline
        \textbf{Conditions} & \textbf{LiDAR point clouds} & \textbf{Communication links} \\
        \hline
        High TTD and ATD    & 45,000   & 337,500   \\ 
        Medium TTD and ATD  & 34,500   & 180,000   \\ 
        Low TTD and ATD     & 16,500   & 36,000    \\ 
        \hline
        \textbf{Total}      & \textbf{96,000} & \textbf{553,500} \\
        \hline
    \end{tabular}
\end{table}
\begin{figure*}[!t]
		\centering
        \subfigure[]{\includegraphics[width=0.32\textwidth,height=0.13\textheight]{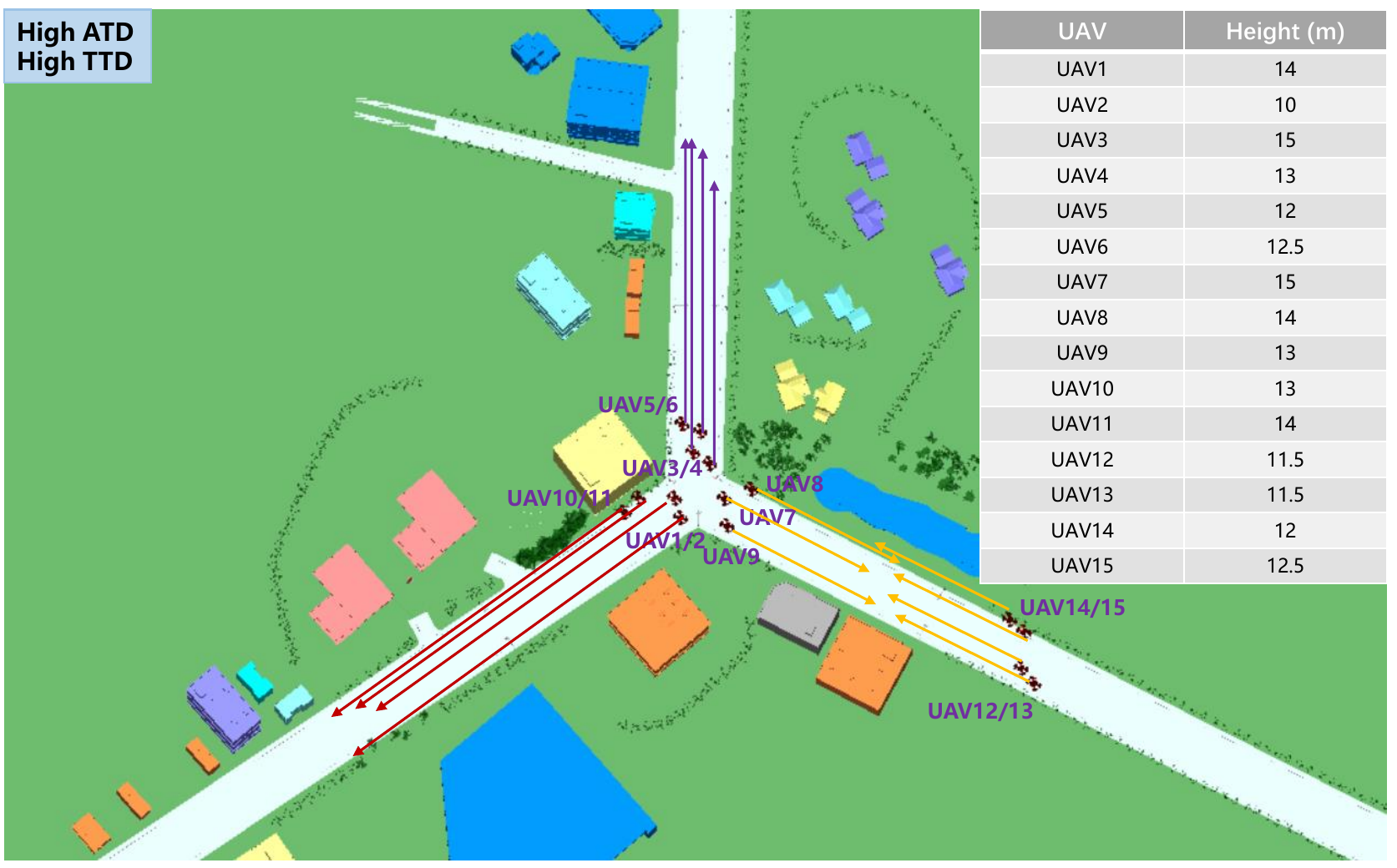}}
        \subfigure[]{\includegraphics[width=0.32\textwidth,height=0.13\textheight]{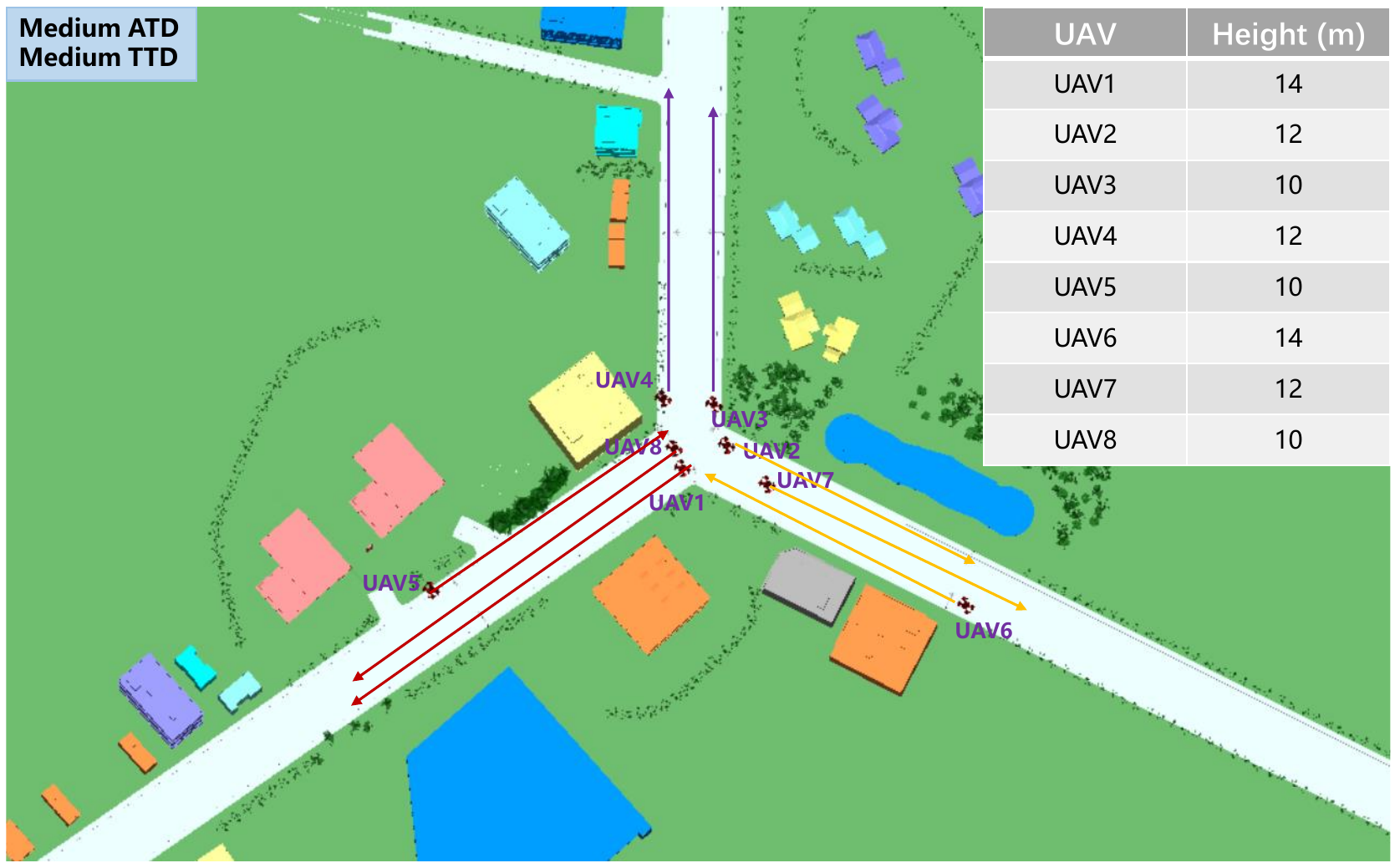}}
        \subfigure[]{\includegraphics[width=0.32\textwidth,height=0.13\textheight]{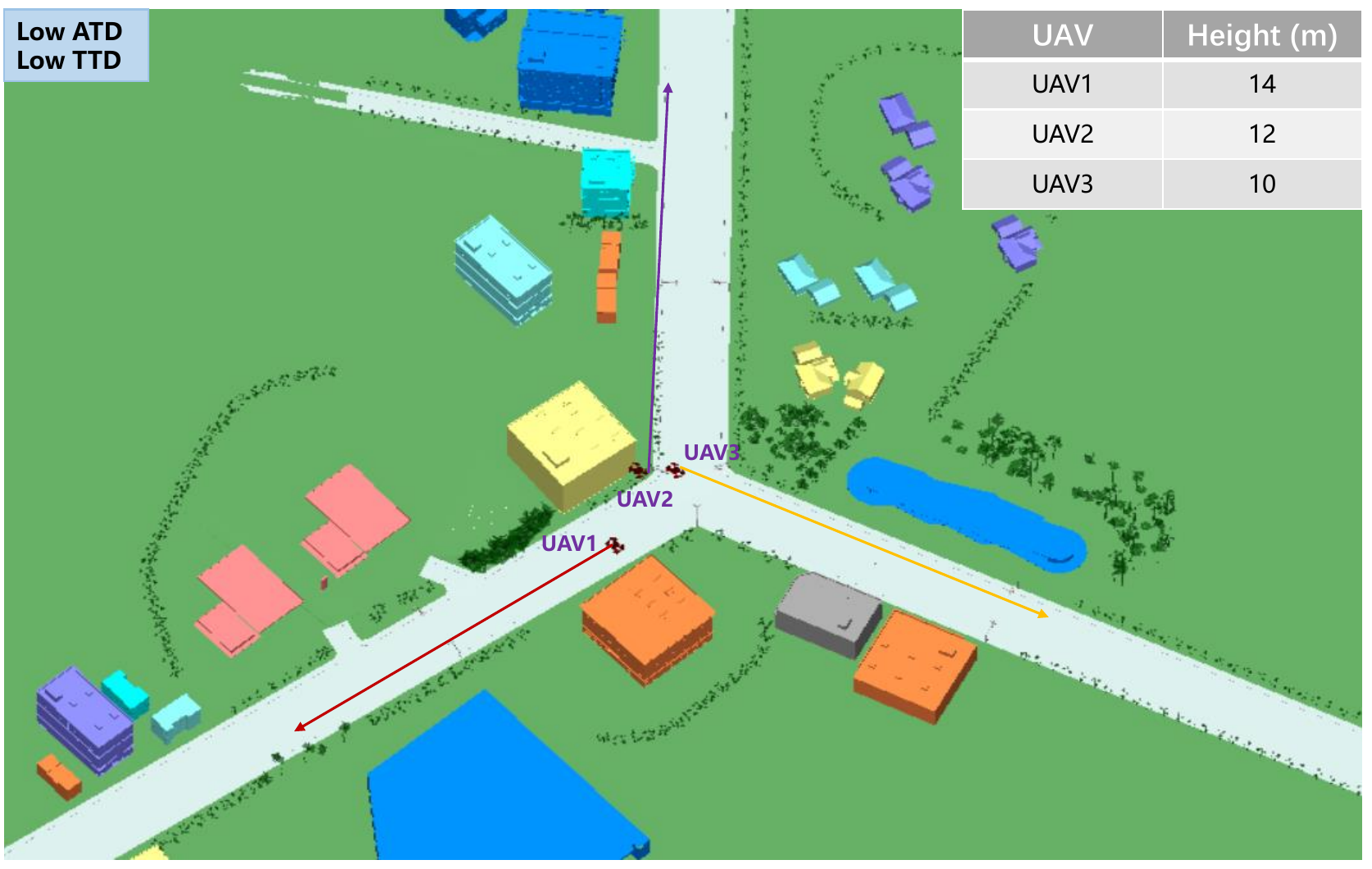}}
	\caption{UAV flight trajectories under multi-UAV-to-multi-vehicle scenarios. Figs. (a)--(c) are the UAV flight trajectories in Wireless InSite under high, medium, and low TTD and ATD conditions, respectively.}
	\label{UAV_trace}
\end{figure*}
\subsection{MUMV-CSCI Dataset and Mapping Relationship Between Electromagnetic Space and Physical Environment in Suburban Forking Road Scenarios}
Since no software can simultaneously collect integrated sensing data and communication data, two simulation platforms, i.e., AirSim \cite{AirSim} and Wireless InSite \cite{WI}, are fused to fulfill the in-depth integration between sensing and communications as well as the precise alignment between the physical environment and electromagnetic space. Each transceiver is equipped with communication equipment and a LiDAR device to collect communication data and sensory data. The carrier frequency of communication equipment is $f_\mathrm{c}=28$~GHz with the bandwidth of $2$~GHz. All of the transceivers are equipped with one antenna. 
To achieve the precise alignment between the physical environment and electromagnetic space, the suburban forking road scenario is constructed in AirSim. Then, it is imported into Wireless InSite to simulate the electromagnetic space. To investigate the impact of traffic density conditions in both terrestrial and aerial areas, the TTD and ATD are developed for the first time. In this paper, the numbers of vehicles under low, medium, and high TTD conditions in the terrestrial areas are 8, 15, and 25, and the numbers of UAVs under low, medium, and high ATD conditions in the aerial areas are 3, 8, and 15. 

The scenarios under high TTD and ATD conditions in AirSim are shown in Fig. \ref{WI_AirSim}. 
In the low TTD and ATD conditions, the communication data of the links between the $1$-st to $3$-rd UAV and the $1$-st to $8$-th vehicles is collected.
In the medium TTD and ATD conditions, the communication data of the links between the $1$-st to $8$-th UAV and the $1$-st to $15$-th vehicles is collected.
In the high TTD and ATD conditions, the communication data of the links between the $1$-st to $15$-th UAV and the $1$-st to $15$-th vehicles is collected.
The flight trajectory and height of UAVs under different TTD and ATD are shown in Fig. \ref{UAV_trace}.
Therefore, a new MUMV-CSCI dataset in the suburban forking road is constructed. For clarity, Table \ref{WI_link} summarizes the data volume size of the sensory data and communication data.

The high mobility of multiple transceivers and scatterers leads to complicated characteristics. Therefore, the detection of dynamic scatterers is of great significance. With the aid of sensing data, i.e., LiDAR point clouds, the static, terrestrial dynamic, and aerial dynamic scatterers are detected and matched with the static, terrestrial dynamic, and aerial dynamic objects.
The raw LiDAR point clouds are redundant and full of useless ground points and should be eliminated. There is only information related to static objects and dynamic objects in the pre-processed LiDAR point clouds, which represent static buildings and facilities as well as dynamic UAVs and vehicles in the physical environment. By exploiting the typical density-based spatial clustering of applications with noise \cite{DBSCAN}, the objects in the physical environment are extracted. According to the sizes of extracted objects, the objects can be classified into static, terrestrial dynamic, and aerial dynamic objects.  
Scatterers obtained from the RT-based wireless channel data coincide with a static/terrestrial dynamic/aerial dynamic object detected from the LiDAR point clouds in the physical environment, the scatterers are determined as a static/terrestrial dynamic/aerial dynamic scatterer in the electromagnetic space.

\begin{table*}[!t]
		\centering
		\caption{\textsc{Key Statistical Parameters in Multi-UAV-to-Multi-Vehicle Sensing and Communication Intelligent Integration Channels}}
		\label{Parameter_1}
				\begin{scriptsize}
		\renewcommand\arraystretch{1}
		\begin{tabular}{|c|c|c|c|c|c|}
			\hline
			\multicolumn{2}{|c|}{Parameter}	& Distribution & Type  & TTD and ATD Conditions & Value  \\
						\hline
      		\multicolumn{2}{|c|}{\multirow{18}{*}{Number}} & \multirow{18}{*}{Logistic}  & \multirow{3}{*}{Static cluster} & High & $\mu^\mathrm{c,L}_\mathrm{s}=0.1511$, $\gamma^\mathrm{c,L}_\mathrm{s}=0.0520$ \\
  \cline{5-6}
  \multicolumn{2}{|c|}{} & & &  Medium & $\mu^\mathrm{c,L}_\mathrm{s}=0.0915$, $\gamma^\mathrm{c,L}_\mathrm{s}=0.0455$\\
    \cline{5-6}
  \multicolumn{2}{|c|}{}& & &  Low & $\mu^\mathrm{c,L}_\mathrm{s}=0.0620$, $\gamma^\mathrm{c,L}_\mathrm{s}=0.0821$\\
    \cline{4-6}
  \multicolumn{2}{|c|}{}& & \multirow{3}{*}{Terrestrial dynamic cluster} & High & $\mu^\mathrm{c,L}_\mathrm{td}=0.1126$, $\gamma^\mathrm{c,L}_\mathrm{td}=0.1015$\\
\cline{5-6}
\multicolumn{2}{|c|}{}& & & Medium & $\mu^\mathrm{c,L}_\mathrm{td}=0.1138$, $\gamma^\mathrm{c,L}_\mathrm{td}=0.0851$\\
    \cline{5-6}
\multicolumn{2}{|c|}{}& & & Low & $\mu^\mathrm{c,L}_\mathrm{td}=0.0842$, $\gamma^\mathrm{c,L}_\mathrm{td}=0.0289$\\
\cline{4-6}
  \multicolumn{2}{|c|}{}& & \multirow{2}{*}{Aerial dynamic cluster} & High & $\mu^\mathrm{c,L}_\mathrm{ad}=0.2356$, $\gamma^\mathrm{c,L}_\mathrm{ad}=0.0321$\\
\cline{5-6}
\multicolumn{2}{|c|}{}& & & Medium & $\mu^\mathrm{c,L}_\mathrm{ad}=0.1825$, $\gamma^\mathrm{c,L}_\mathrm{ad}=0.0528$\\
    \cline{5-6}
  \cline{4-6}
 \multicolumn{2}{|c|}{} & & \multirow{3}{*}{Static scatterer} & High & $\mu^\mathrm{s,L}_\mathrm{s}=0.7534$, $\gamma^\mathrm{s,L}_\mathrm{s}=0.5236$\\
  \cline{5-6}
\multicolumn{2}{|c|}{}  & & & Medium & $\mu^\mathrm{s,L}_\mathrm{s}=0.6024$, $\gamma^\mathrm{s,L}_\mathrm{s}=0.4726$\\
    \cline{5-6}
\multicolumn{2}{|c|}{}  & & & Low & $\mu^\mathrm{s,L}_\mathrm{s}=0.3425$, $\gamma^\mathrm{s,L}_\mathrm{s}=0.3855$\\
    \cline{4-6}
\multicolumn{2}{|c|}{}    & & \multirow{3}{*}{Terrestrial dynamic scatterer} & High & $\mu^\mathrm{s,L}_\mathrm{td}=0.4461$, $\gamma^\mathrm{s,L}_\mathrm{td}=0.3921$\\
\cline{5-6}
\multicolumn{2}{|c|}{}& & & Medium & $\mu^\mathrm{s,L}_\mathrm{td}=0.3928$, $\gamma^\mathrm{s,L}_\mathrm{td}=0.2511$\\
    \cline{5-6}
\multicolumn{2}{|c|}{}& & & Low & $\mu^\mathrm{s,L}_\mathrm{td}=0.3213$, $\gamma^\mathrm{s,L}_\mathrm{td}=0.1863$\\
    \cline{4-6}
\multicolumn{2}{|c|}{}    & & \multirow{2}{*}{Aerial dynamic scatterer} & High & $\mu^\mathrm{s,L}_\mathrm{ad}=0.4232$, $\gamma^\mathrm{s,L}_\mathrm{ad}=0.4261$\\
\cline{5-6}
\multicolumn{2}{|c|}{}& & & Medium & $\mu^\mathrm{s,L}_\mathrm{ad}=0.3821$, $\gamma^\mathrm{s,L}_\mathrm{ad}=0.3925$\\
    \cline{5-6}
    \hline
\multicolumn{2}{|c|}{\multirow{9}{*}{Distance}}  & \multirow{3}{*}{Gamma}  & \multirow{3}{*}{Static scatterer} & High & $\alpha^\mathrm{G}_\mathrm{s}=0.8223$, $\beta^\mathrm{G}_\mathrm{s}=1.9232$ \\
  \cline{5-6}
\multicolumn{2}{|c|}{} & & &  Medium & $\alpha^\mathrm{G}_\mathrm{s}=0.6982$, $\beta^\mathrm{G}_\mathrm{s}=2.0263$\\
    \cline{5-6}
\multicolumn{2}{|c|}{}  & & &  Low & $\alpha^\mathrm{G}_\mathrm{s}=0.6241$, $\beta^\mathrm{G}_\mathrm{s}=2.4581$\\
    \cline{3-6}
\multicolumn{2}{|c|}{}& \multirow{3}{*}{Rayleigh}  & \multirow{3}{*}{Terrestrial dynamic scatterer} & High & $\sigma^\mathrm{R}_\mathrm{td}=0.3541$ \\
  \cline{5-6}
 \multicolumn{2}{|c|}{} & & &  Medium & $\sigma^\mathrm{R}_\mathrm{td}=0.3026$\\
    \cline{5-6}
  \multicolumn{2}{|c|}{} & & &  Low & $\sigma^\mathrm{R}_\mathrm{td}=0.2025$\\
     \cline{3-6}
  \multicolumn{2}{|c|}{}& \multirow{2}{*}{Rayleigh}  & \multirow{2}{*}{Aerial dynamic scatterer} & High & $\sigma^\mathrm{R}_\mathrm{ad}=0.3356$ \\
  \cline{5-6}
 \multicolumn{2}{|c|}{} & & &  Medium & $\sigma^\mathrm{R}_\mathrm{ad}=0.2287$\\
    \cline{5-6}
    \hline  
      \multirow{36}{*}{Angle} & \multirow{9}{*}{AAoD} & \multirow{9}{*}{Gaussian}  & \multirow{3}{*}{Static scatterer} & High & $\mu^\mathrm{AAoD}_\mathrm{s}=0.8254$, $\sigma^\mathrm{AAoD}_\mathrm{s}=0.9254$ \\
  \cline{5-6}
  & & & &  Medium & $\mu^\mathrm{AAoD}_\mathrm{s}=0.7612$, $\sigma^\mathrm{AAoD}_\mathrm{s}=0.8723$\\
    \cline{5-6}
  & & & &  Low & $\mu^\mathrm{AAoD}_\mathrm{s}=0.7025$, $\sigma^\mathrm{AAoD}_\mathrm{s}=0.7566$\\
        \cline{4-6} 
       &  &   & \multirow{3}{*}{Terrestrial dynamic scatterer} & High & $\mu^\mathrm{AAoD}_\mathrm{td}=0.9213$, $\sigma^\mathrm{AAoD}_\mathrm{td}=1.9253$ \\
  \cline{5-6}
  & & & &  Medium & $\mu^\mathrm{AAoD}_\mathrm{td}=0.8190$, $\sigma^\mathrm{AAoD}_\mathrm{td}=1.7622$\\
    \cline{5-6}
  & & & &  Low & $\mu^\mathrm{AAoD}_\mathrm{td}=0.7623$, $\sigma^\mathrm{AAoD}_\mathrm{td}=1.2101$\\
   \cline{4-6} 
       &  &   & \multirow{2}{*}{Aerial dynamic scatterer} & High & $\mu^\mathrm{AAoD}_\mathrm{ad}=0.3241$, $\sigma^\mathrm{AAoD}_\mathrm{ad}=1.0125$ \\
  \cline{5-6}
  & & & &  Medium & $\mu^\mathrm{AAoD}_\mathrm{ad}=0.2015$, $\sigma^\mathrm{AAoD}_\mathrm{ad}=0.9215$\\
    \cline{5-6}
    \cline{2-3}      \cline{4-6} 
     & \multirow{9}{*}{AAoA} & \multirow{9}{*}{Gaussian}  & \multirow{3}{*}{Static scatterer} & High & $\mu^\mathrm{AAoA}_\mathrm{s}=0.4521$, $\sigma^\mathrm{AAoA}_\mathrm{s}=0.4834$ \\
  \cline{5-6}
  & & & &  Medium & $\mu^\mathrm{AAoA}_\mathrm{s}=0.4025$, $\sigma^\mathrm{AAoA}_\mathrm{s}=0.4512$\\
    \cline{5-6}
  & & & &  Low & $\mu^\mathrm{AAoA}_\mathrm{s}=0.3816$, $\sigma^\mathrm{AAoA}_\mathrm{s}=0.3266$\\
       \cline{4-6}
     &  &  & \multirow{3}{*}{Terrestral dynamic scatterer} & High & $\mu^\mathrm{AAoA}_\mathrm{td}=-0.3215$, $\sigma^\mathrm{AAoA}_\mathrm{td}=0.5124$ \\
  \cline{5-6}
  & & & &  Medium & $\mu^\mathrm{AAoA}_\mathrm{td}=-0.4156$, $\sigma^\mathrm{AAoA}_\mathrm{td}=0.4266$\\
    \cline{5-6}
  & & & &  Low & $\mu^\mathrm{AAoA}_\mathrm{td}=-0.2511$, $\sigma^\mathrm{AAoA}_\mathrm{td}=0.1756$\\
   \cline{4-6}
     &  &  & \multirow{2}{*}{Aerial dynamic scatterer} & High & $\mu^\mathrm{AAoA}_\mathrm{ad}=0.5416$, $\sigma^\mathrm{AAoA}_\mathrm{ad}=0.6524$ \\
  \cline{5-6}
  & & & &  Medium & $\mu^\mathrm{AAoA}_\mathrm{ad}=0.4211$, $\sigma^\mathrm{AAoA}_\mathrm{ad}=0.5815$\\
    \cline{5-6}
   \cline{2-3} \cline{4-6}
     & \multirow{9}{*}{EAoD} & \multirow{9}{*}{Gaussian} & \multirow{3}{*}{Static scatterer} & High & $\mu^\mathrm{EAoD}_\mathrm{s}=0.7514$, $\sigma^\mathrm{EAoD}_\mathrm{s}=0.8512$ \\
  \cline{5-6}
  & & & &  Medium & $\mu^\mathrm{EAoD}_\mathrm{s}=0.7142$, $\sigma^\mathrm{EAoD}_\mathrm{s}=0.6215$\\
    \cline{5-6}
  & & & &  Low & $\mu^\mathrm{EAoD}_\mathrm{s}=0.7836$, $\sigma^\mathrm{EAoD}_\mathrm{s}=0.4315$\\
       \cline{4-6}
     &  &  & \multirow{3}{*}{Terrestrial dynamic scatterer} & High & $\mu^\mathrm{EAoD}_\mathrm{td}=0.1545$, $\sigma^\mathrm{EAoD}_\mathrm{td}=0.7851$ \\
  \cline{5-6}
  & & & &  Medium & $\mu^\mathrm{EAoD}_\mathrm{td}=0.1951$, $\sigma^\mathrm{EAoD}_\mathrm{td}=0.7011$\\
    \cline{5-6}
  & & & &  Low & $\mu^\mathrm{EAoD}_\mathrm{td}=0.1766$, $\sigma^\mathrm{EAoD}_\mathrm{td}=0.6789$\\
     \cline{4-6}
     &  &  & \multirow{2}{*}{Aerial dynamic scatterer} & High & $\mu^\mathrm{EAoD}_\mathrm{ad}=0.9511$, $\sigma^\mathrm{EAoD}_\mathrm{ad}=1.8251$ \\
  \cline{5-6}
  & & & &  Medium & $\mu^\mathrm{EAoD}_\mathrm{ad}=0.9151$, $\sigma^\mathrm{EAoD}_\mathrm{ad}=1.6435$\\
    \cline{5-6}
\cline{2-3} \cline{4-6}
     & \multirow{9}{*}{EAoA} &\multirow{9}{*}{Gaussian}  & \multirow{3}{*}{Static scatterer} & High & $\mu^\mathrm{EAoA}_\mathrm{s}=0.8516$, $\sigma^\mathrm{EAoA}_\mathrm{s}=0.7612$ \\
  \cline{5-6}
  & & & &  Medium & $\mu^\mathrm{EAoA}_\mathrm{s}=0.8781$, $\sigma^\mathrm{EAoA}_\mathrm{s}=0.6921$\\
    \cline{5-6}
  & & & &  Low & $\mu^\mathrm{EAoA}_\mathrm{s}=0.8423$, $\sigma^\mathrm{EAoA}_\mathrm{s}=0.5516$\\
       \cline{4-6}
     &  &  & \multirow{3}{*}{Terrestrial dynamic scatterer} & High & $\mu^\mathrm{EAoA}_\mathrm{td}=0.2511$, $\sigma^\mathrm{EAoA}_\mathrm{td}=0.9218$ \\
  \cline{5-6}
  & & & &  Medium & $\mu^\mathrm{EAoA}_\mathrm{td}=0.1921$, $\sigma^\mathrm{EAoA}_\mathrm{td}=0.9055$\\
    \cline{5-6}
  & & & &  Low & $\mu^\mathrm{EAoA}_\mathrm{td}=0.2249$, $\sigma^\mathrm{EAoA}_\mathrm{td}=0.8127$\\
   \cline{4-6}
     &  &  & \multirow{2}{*}{Aerial dynamic scatterer} & High & $\mu^\mathrm{EAoA}_\mathrm{ad}=0.8915$, $\sigma^\mathrm{EAoA}_\mathrm{ad}=1.9627$ \\
  \cline{5-6}
  & & & &  Medium & $\mu^\mathrm{EAoA}_\mathrm{ad}=0.7812$, $\sigma^\mathrm{EAoA}_\mathrm{ad}=1.8541$\\
    \cline{5-6}
   \hline
\multicolumn{2}{|c|}{\multirow{9}{*}{Power-Delay}}  & \multirow{9}{*}{Exponential}  & \multirow{3}{*}{Static scatterer} & High & $\xi_\mathrm{s}=2.6881\times10^6$, $\eta_\mathrm{s}=31.9204$, $\sigma_\mathrm{E,s}=19.9350$ \\
  \cline{5-6}
\multicolumn{2}{|c|}{} & & &  Medium & $\xi_\mathrm{s}=4.8043\times10^6$, $\eta_\mathrm{s}=30.4251$, $\sigma_\mathrm{E,s}=22.3581$ \\
    \cline{5-6}
\multicolumn{2}{|c|}{}  & & &  Low & $\xi_\mathrm{s}=2.2978\times10^6$, $\eta_\mathrm{s}=30.0112$, $\sigma_\mathrm{E,s}=16.1603$\\
    \cline{4-6}
\multicolumn{2}{|c|}{}&  & \multirow{3}{*}{Terrestrial dynamic scatterer} & High & $\xi_\mathrm{td}=2.1931\times10^6$, $\eta_\mathrm{td}=31.3934$, $\sigma_\mathrm{E,td}=11.6472$ \\
  \cline{5-6}
 \multicolumn{2}{|c|}{} & & &  Medium & $\xi_\mathrm{td}=3.6554\times10^6$, $\eta_\mathrm{td}=30.5136$, $\sigma_\mathrm{E,td}=13.6758$ \\
    \cline{5-6}
  \multicolumn{2}{|c|}{} & & &  Low & $\xi_\mathrm{td}=1.2030\times10^6$, $\eta_\mathrm{td}=31.4610$, $\sigma_\mathrm{E,td}=0.2222$ \\
      \cline{4-6}
\multicolumn{2}{|c|}{}&  & \multirow{2}{*}{Aerial dynamic scatterer} & High & $\xi_\mathrm{ad}=3.9797\times10^6$, $\eta_\mathrm{ad}=29.2900$, $\sigma_\mathrm{E,ad}=12.0014$ \\
  \cline{5-6}
 \multicolumn{2}{|c|}{} & & &  Medium & $\xi_\mathrm{ad}=5.5346\times10^6$, $\eta_\mathrm{ad}=28.5798$, $\sigma_\mathrm{E,ad}=9.8293$ \\
    \cline{5-6}
    \hline  
		\end{tabular}
			\end{scriptsize}
	\end{table*}

\subsection{Channel Parameterization and Characterization}
\subsubsection{Numbers of Scatterers and Clusters}
\begin{figure*}[!t]
	\centering
	\subfigure[]{\includegraphics[width=0.32\textwidth]{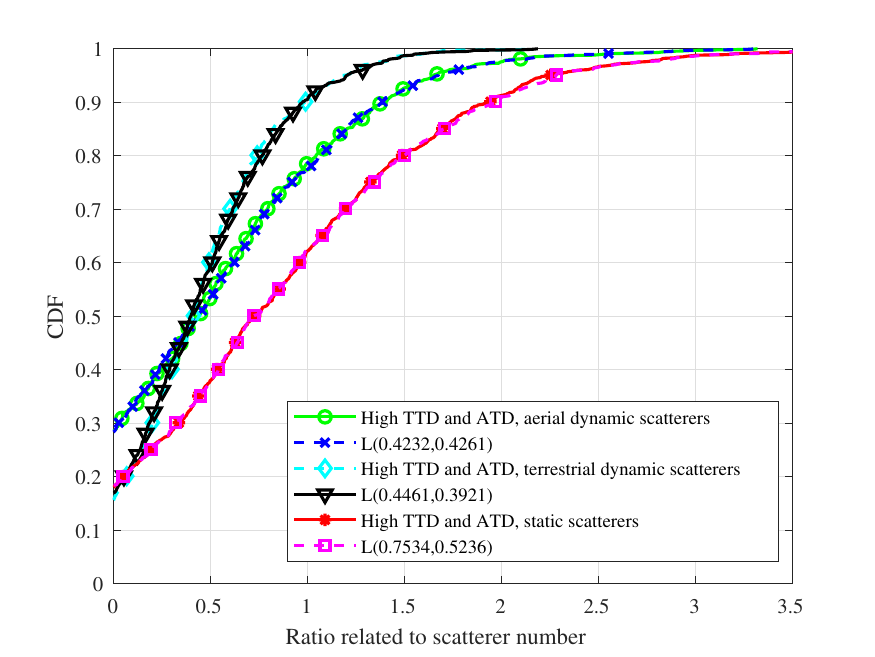}}
	\subfigure[]{\includegraphics[width=0.32\textwidth]{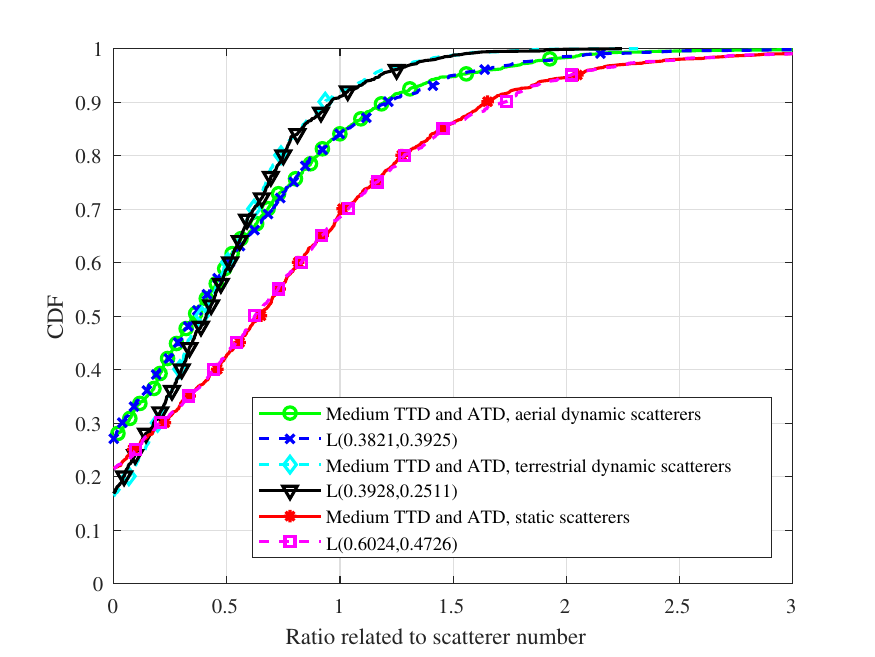}} 
   \subfigure[]{\includegraphics[width=0.32\textwidth]{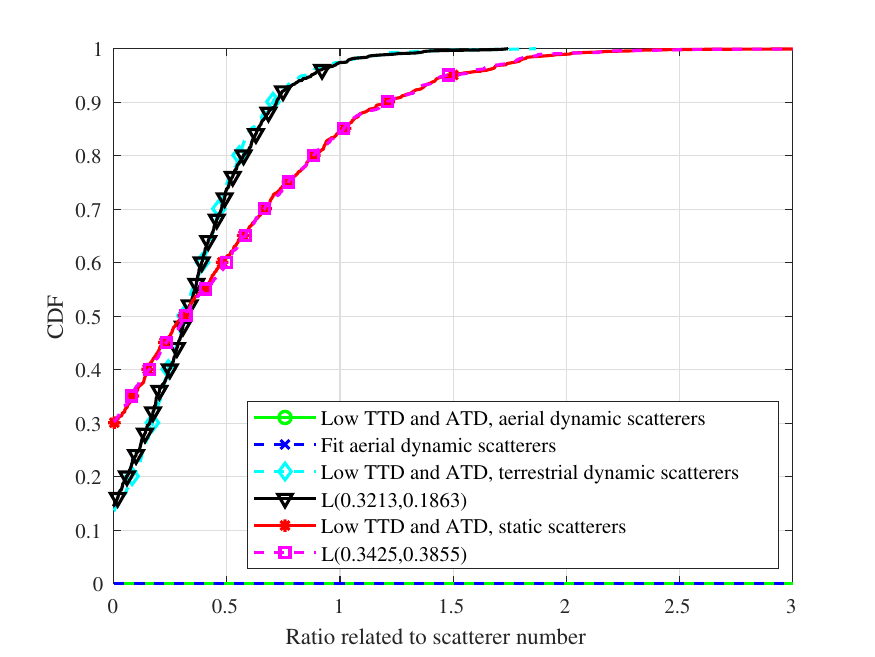}}
    \subfigure[]{\includegraphics[width=0.32\textwidth]{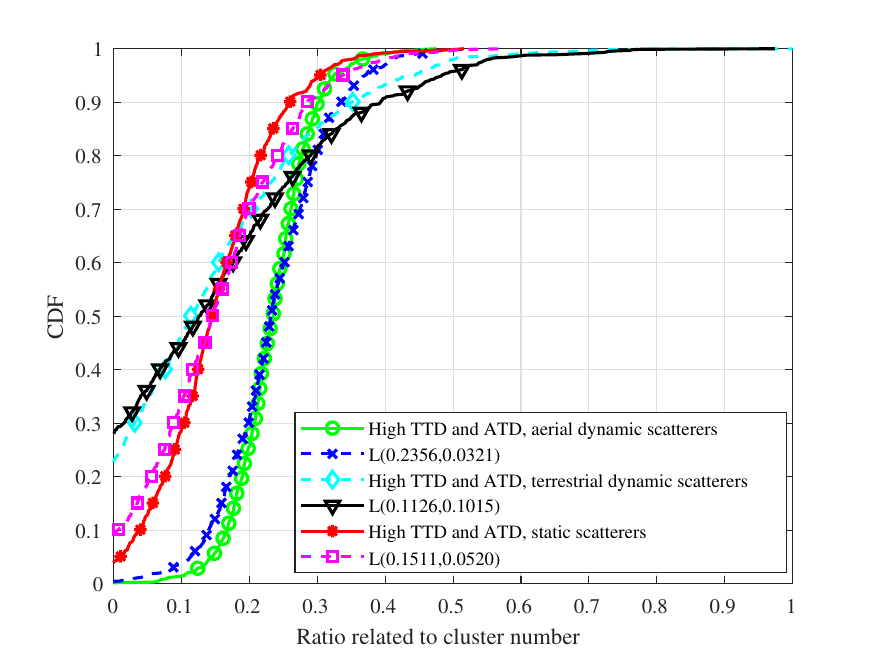}}  \subfigure[]{\includegraphics[width=0.32\textwidth]{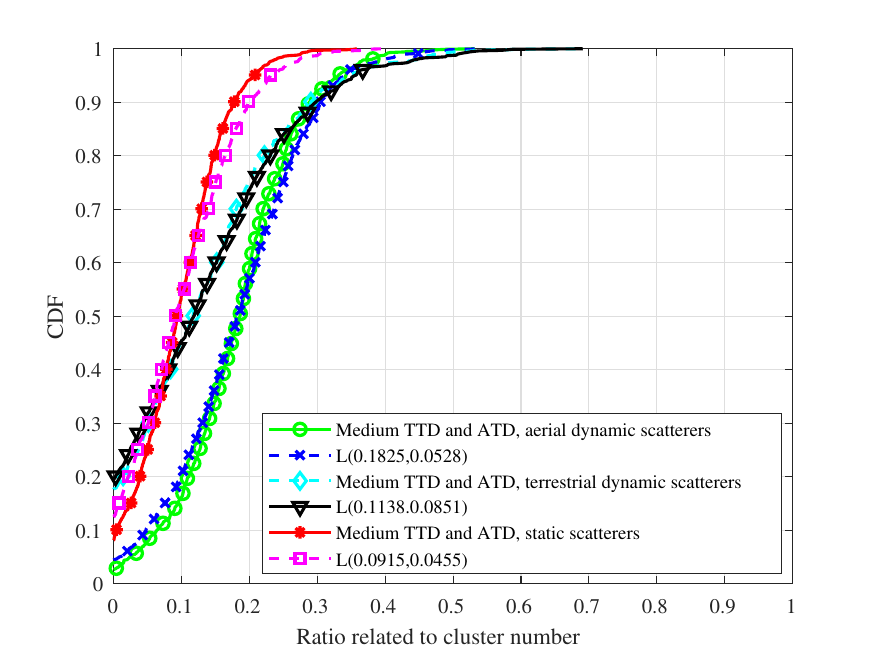}}
\subfigure[]{\includegraphics[width=0.32\textwidth]{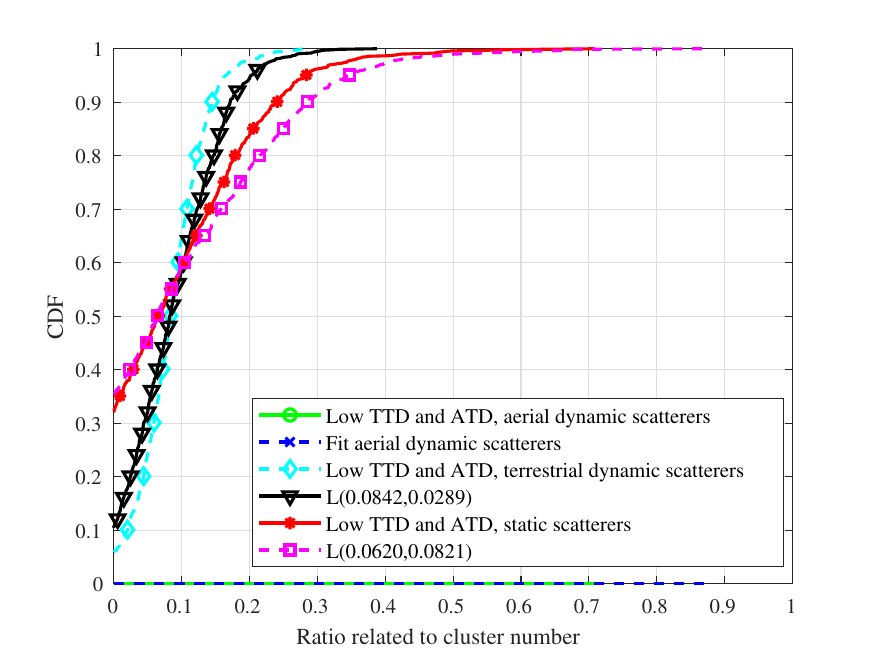}}
	\caption{CDFs of static/terrestrial dynamic/aerial dynamic scatterer and cluster number parameters with the Logistic distribution fitting under different TTD and ATD conditions. Figs.~(a)--(c) show the CDFs of scatterer number parameters under high, medium, and low TTD and ATD conditions, respectively. Figs.~(d)--(f) show the CDFs of cluster number parameters under high, medium, and low TTD and ATD conditions, respectively.}
 \label{Number}
\end{figure*}

Currently, the numbers of static, terrestrial dynamic, and aerial dynamic scatterers in standardized models \cite{METIS}--\cite{Qua} are not differentiated modeling. Meanwhile, cooperation communications between multi-UAVs and multi-vehicles are also not considered. With the aid of LiDAR point clouds, static, terrestrial dynamic, and aerial dynamic scatterers can be accurately distinguished.
The numbers of static, terrestrial dynamic, aerial dynamic scatterers in the transmission link from the $i$-th UAV ($i=1, 2, ..., I$), i.e., transmitter (Tx), to the $j$-th vehicle  ($j=1, 2, ..., J$), i.e., receiver (Rx), are denoted as $B^{\mathrm{U}_i,\mathrm{C}_j}_\mathrm{s}(t)$, $B^{\mathrm{U}_i,\mathrm{C}_j}_\mathrm{td}(t)$, and $B^{\mathrm{U}_i,\mathrm{C}_j}_\mathrm{ad}(t)$. Considering the impact of transmission distance, the static, terrestrial dynamic, aerial dynamic scatterer ratios, i.e., $N^{\mathrm{U}_i,\mathrm{C}_j}_\mathrm{s}(t)$, $N^{\mathrm{U}_i,\mathrm{C}_j}_\mathrm{td}(t)$, and $N^{\mathrm{U}_i,\mathrm{C}_j}_\mathrm{ad}(t)$, are introduced. The static/terrestrial dynamic/aerial dynamic scatterer number ratio represents the ratio of static/terrestrial dynamic/aerial dynamic scatterer number to the distance between the $i$-th UAV and the $j$-th vehicle, which is calculated by
 \begin{equation}
     N^{\mathrm{U}_i,\mathrm{C}_j}_\mathrm{s}(t)=\frac{B^{\mathrm{U}_i,\mathrm{C}_j}_\mathrm{s}(t)}{\|\mathbf{T}^{\mathrm{U}_i}(t)-\mathbf{R}^{\mathrm{C}_j}(t)\|}
 \end{equation}
 \begin{equation}
     N^{\mathrm{U}_i,\mathrm{C}_j}_\mathrm{td}(t)=\frac{B^{\mathrm{U}_i,\mathrm{C}_j}_\mathrm{td}(t)}{\|\mathbf{T}^{\mathrm{U}_i}(t)-\mathbf{R}^{\mathrm{C}_j}(t)\|}
 \end{equation}
  \begin{equation}
     N^{\mathrm{U}_i,\mathrm{C}_j}_\mathrm{ad}(t)=\frac{B^{\mathrm{U}_i,\mathrm{C}_j}_\mathrm{ad}(t)}{\|\mathbf{T}^{\mathrm{U}_i}(t)-\mathbf{R}^{\mathrm{C}_j}(t)\|}
 \end{equation}
 where $\mathbf{T}^{\mathrm{U}_i}(t)$ and $\mathbf{R}^{\mathrm{C}_j}(t)$ are the locations of the $i$-th UAV and the $j$-th vehicle. Moreover, based on the constructed MUMV-CSCI dataset, the static, terrestrial dynamic, and aerial dynamic scatterer number ratios in each communication link at each snapshot are calculated and analyzed. Figs.~\ref{Number}(a)--(c) gives the cumulative distribution functions (CDFs) of static, terrestrial dynamic, and aerial dynamic scatterer number ratios under low, medium, and high TTD and ATD, respectively. 
 The CDF of static/terrestrial dynamic/aerial dynamic scatterer number ratio fits well with the Logistic distribution, which is given by
 \begin{equation}
     F^\mathrm{s,L}_\mathrm{s/td/ad}(x)=\frac{1}{1+e^{-(x-\mu^\mathrm{s,L}_\mathrm{s/td/ad})/{\gamma^\mathrm{s,L}_\mathrm{s/td/ad}}}}
 \end{equation}
 where $\mu^\mathrm{s,L}_\mathrm{s/td/ad}$ and $\gamma^\mathrm{s,L}_\mathrm{s/td/ad}$ are the mean value and the scale parameter of the Logistic distribution for static/terrestrial dynamic/aerial dynamic scatterers.
From Table~\ref{Parameter_1} and Fig.~\ref{Number}, it can be seen that as the TTD and ATD conditions increase, the mean value and variance value of the Logistic distribution for the CDF of dynamic scatterers increase. This phenomenon can be explained that the number of dynamic scatterers increases as the number of dynamic vehicles and UAVs around the transceiver increases. 
To further investigate channel characteristics, the static/terrestrial dynamic/aerial dynamic scatterers are clustered to explore the statistical distribution of the numbers of static/terrestrial dynamic/aerial dynamic clusters. Three new cluster number parameters, $M^{\mathrm{U}_i,\mathrm{C}_j}_\mathrm{s}(t)$, $M^{\mathrm{U}_i,\mathrm{C}_j}_\mathrm{td}(t)$, and $M^{\mathrm{U}_i,\mathrm{C}_j}_\mathrm{ad}(t)$, which represent the ratios of static, terrestrial dynamic, and aerial dynamic cluster numbers to the distance between the $i$-th UAV and the $j$-th vehicle, are introduced. The number of parameters of static, terrestrial dynamic, and aerial dynamic clusters for each communication link at each snapshot is calculated.
Figs.~\ref{Number}(d)--(f) illustrate the CDFs of static, terrestrial dynamic, and aerial dynamic clusters under high, medium, and low TTD and ATD conditions. The Logistic distribution for the CDF of static/terrestrial dynamic/aerial dynamic scatterer clusters can be represented as
 \begin{equation}
     F^\mathrm{c,L}_\mathrm{s/td/ad}(x)=\frac{1}{1+e^{-(x-\mu^\mathrm{c,L}_\mathrm{s/td/ad})/{\gamma^\mathrm{c,L}_\mathrm{s/td/ad}}}}
 \end{equation}
 where $\mu^\mathrm{c,L}_\mathrm{s/td/ad}$ and $\gamma^\mathrm{c,L}_\mathrm{s/td/ad}$ are the mean value and scale parameter of the Logistic distribution for static/terrestrial dynamic/aerial dynamic clusters.
From Table~\ref{Parameter_1} and Fig.~\ref{Number}, it can be seen that the observation of static/terrestrial dynamic/aerial dynamic cluster number parameters is similar to that of static/terrestrial dynamic/aerial dynamic scatterer number parameters.
\begin{figure*}[!t]
	\centering
	\subfigure[]{\includegraphics[width=0.32\textwidth]{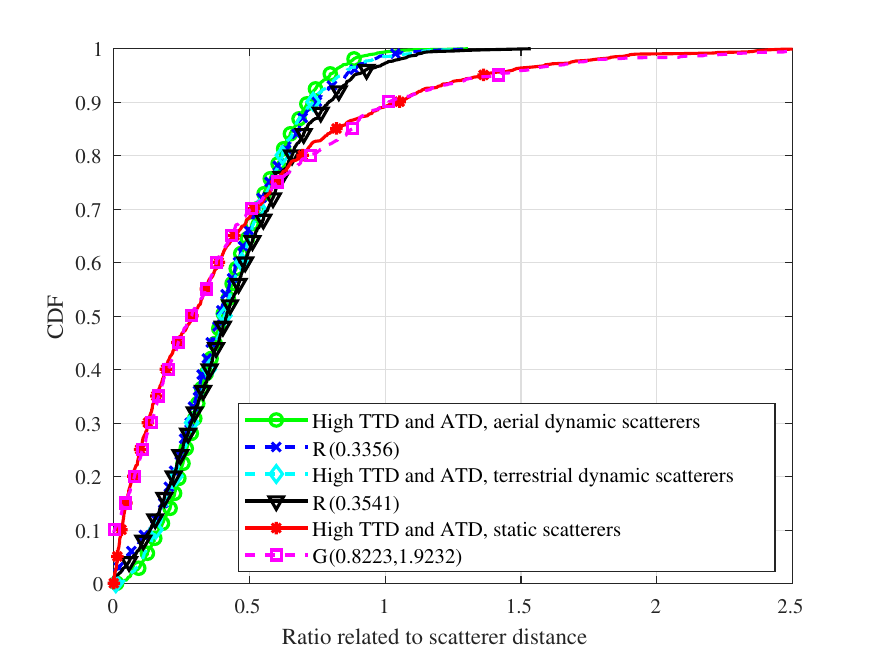}}
	\subfigure[]{\includegraphics[width=0.32\textwidth]{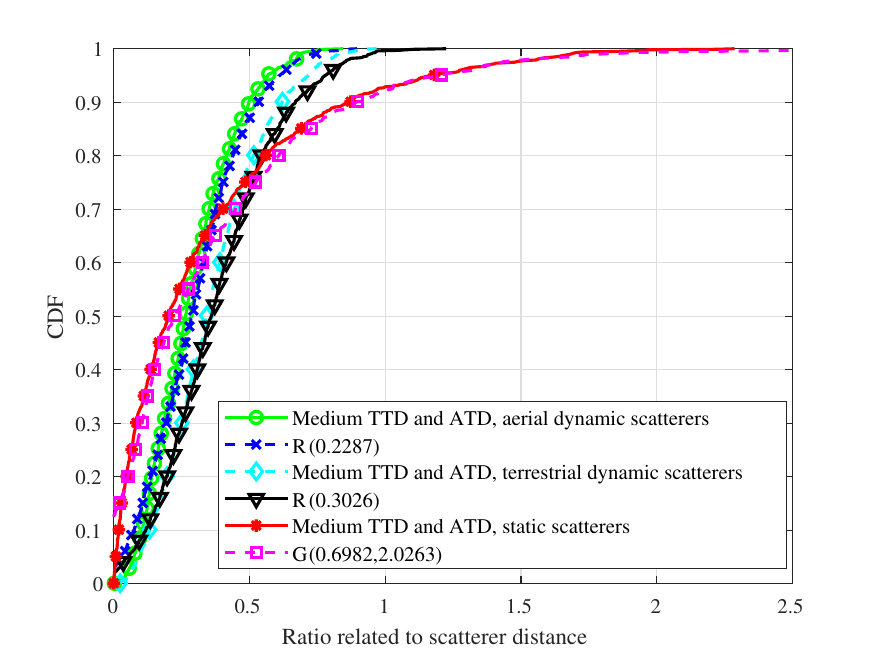}} 
   \subfigure[]{\includegraphics[width=0.32\textwidth]{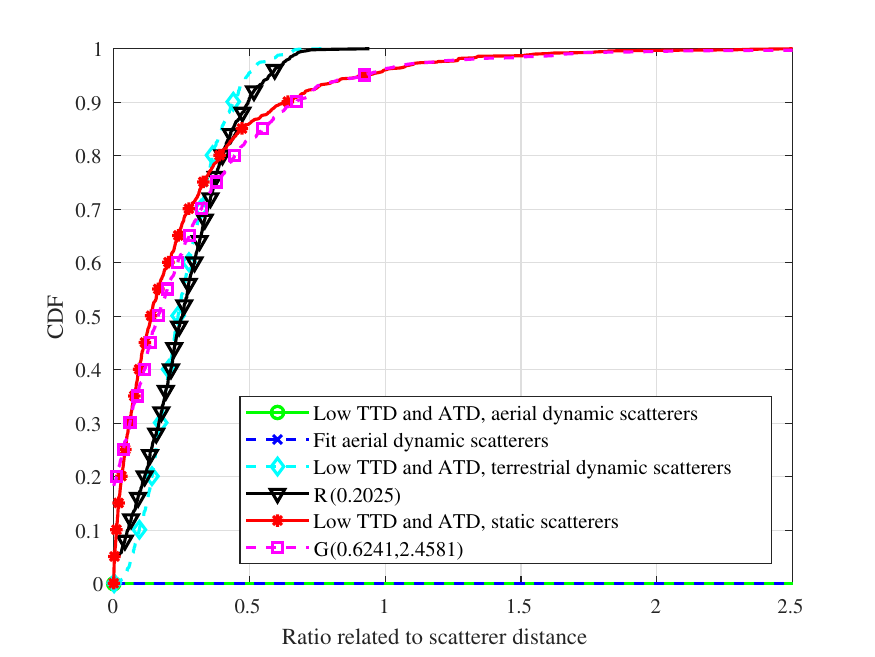}}
	\caption{
    CDFs of static/terrestrial dynamic/aerial dynamic scatterer distance parameters with the Gamma/Rayleigh distribution fitting under different TTD and ATD conditions.  Figs.~(a)--(c) show the CDFs of distance parameters under high, medium, and low TTD and ATD conditions, respectively.}
 \label{Distance}
\end{figure*}
\subsubsection{Distance Parameters}
At present, there is no channel measurement or channel model considering the distinction among distance parameters of static, terrestrial dynamic, and aerial dynamic scatterers/clusters. 
Based on the constructed MUMV-CSCI dataset, the distance parameters of static, terrestrial dynamic, and aerial dynamic scatterers in multi-UAV-to-multi-vehicle channels are analyzed under high, medium, and low TTD and ATD conditions.
The distance parameters from the Txs, i.e., the $i$-th UAV and the $j$-th vehicle, to the $l/m/n$-th static/terrestrial dynamic/aerial dynamic scatterer, i.e., $D^{\mathrm{U}_i,\mathrm{C}_j}_{\mathrm{S}_l}(t)$, $D^{\mathrm{U}_i,\mathrm{C}_j}_{\mathrm{TD}_m}(t)$, and $D^{\mathrm{U}_i,\mathrm{C}_j}_{\mathrm{AD}_n}$, are introduced and expressed as \eqref{distance1}--\eqref{distance3},
\begin{figure*}
    \begin{equation}
\begin{aligned}
 & D^{\mathrm{U}_i,\mathrm{C}_j}_{\mathrm{S}_l}(t)= \frac{\|\mathbf{T}^{\mathrm{U}_i}(t)-\mathbf{S}^{\mathrm{U}_i,\mathrm{C}_j}_{\mathrm{S}_l}(t)\|+\|\mathbf{R}^{\mathrm{C}_j}(t)- 
  \mathbf{S}^{\mathrm{U}_i,\mathrm{C}_j}_{\mathrm{S}_l}(t)\|-\|\mathbf{T}^{\mathrm{U}_i}(t)-\mathbf{R}^{\mathrm{C}_j}(t)\|}{\|\mathbf{T}^{\mathrm{U}_i}(t)-\mathbf{R}^{\mathrm{C}_j}(t)\|}
  \end{aligned}
  \label{distance1}
\end{equation}
\begin{equation}
\begin{aligned}
 & D^{\mathrm{U}_i,\mathrm{C}_j}_{\mathrm{TD}_m}(t)= \frac{\|\mathbf{T}^{\mathrm{U}_i}(t)-\mathbf{S}^{\mathrm{U}_i,\mathrm{C}_j}_{\mathrm{TD}_m}(t)\|+\|\mathbf{R}^{\mathrm{C}_j}(t)-\mathbf{S}^{\mathrm{U}_i,\mathrm{C}_j}_{\mathrm{TD}_m}(t)\|-\|\mathbf{T}^{\mathrm{U}_i}(t)-\mathbf{R}^{\mathrm{C}_j}(t)\|}{\|\mathbf{T}^{\mathrm{U}_i}(t)-\mathbf{R}^{\mathrm{C}_j}(t)\|}
  \end{aligned}
  \label{distance2}
\end{equation}
\begin{equation}
\begin{aligned}
 & D^{\mathrm{U}_i,\mathrm{C}_j}_{\mathrm{AD}_n}(t)= \frac{\|\mathbf{T}^{\mathrm{U}_i}(t)-\mathbf{S}^{\mathrm{U}_i,\mathrm{C}_j}_{\mathrm{AD}_n}(t)\|+\|\mathbf{R}^{\mathrm{C}_j}(t)-\mathbf{S}^{\mathrm{U}_i,\mathrm{C}_j}_{\mathrm{AD}_n}(t)\|-\|\mathbf{T}^{\mathrm{U}_i}(t)-\mathbf{R}^{\mathrm{C}_j}(t)\|}{\|\mathbf{T}^{\mathrm{U}_i}(t)-\mathbf{R}^{\mathrm{C}_j}(t)\|}
  \end{aligned}
  \label{distance3}
\end{equation}
\hrulefill
\vspace*{4pt}
\end{figure*}where $\mathbf{S}^{\mathrm{U}_i,\mathrm{C}_j}_{\mathrm{S}_l}(t)/\mathbf{S}^{\mathrm{U}_i,\mathrm{C}_j}_{\mathrm{TD}_m}(t)/\mathbf{S}^{\mathrm{U}_i,\mathrm{C}_j}_{\mathrm{AD}_n}(t)$ is the location of the $l/m/n$-th static/terrestrial dynamic/aerial dynamic scatterer in the transmission link between the $i$-th UAV and the $j$-th vehicle. $\|\cdot\|$ denotes the calculation of the Frobenius norm. 
Moreover, based on the constructed MUMV-CSCI dataset, the distance parameter of each static/terrestrial dynamic/aerial dynamic scatterer is calculated and analyzed.
Figs.~\ref{Distance}(a)--(c) show the CDFs of distance parameters of static, terrestrial dynamic, and aerial dynamic scatterers under high, medium, and low TTD and ATD conditions, respectively. 
The CDFs of distance parameters of static, terrestrial dynamic, and aerial dynamic scatterer match well with the Gamma distribution, Rayleigh distribution, and Rayleigh distribution, respectively. The CDFs of the
Gamma distribution and the Rayleigh distribution are represented as 
\begin{equation}
F^\mathrm{G}_\mathrm{s}(x)=\frac{\gamma(\alpha^\mathrm{G}_\mathrm{s}, \beta^\mathrm{G}_\mathrm{s} x)}{\Gamma(\alpha^\mathrm{G}_\mathrm{s})}
\end{equation}
\begin{equation}
F^\mathrm{R}_\mathrm{td/ad}(x)=1-e^{-\frac{x^2}{2 (\sigma^\mathrm{R}_\mathrm{td/ad})^2}}
\end{equation}
where $\alpha^\mathrm{G}_\mathrm{s}$ and $\beta^\mathrm{G}_\mathrm{s}$ denote the shape parameter and the rate parameter of Gamma distribution. 
$\Gamma(\cdot)$ and $\gamma(\cdot,\cdot)$ denote the Gamma function and the lower incomplete Gamma function. $\sigma^\mathrm{R}_\mathrm{td/ad}$ denotes the scale parameter of Rayleigh distribution. 
As shown in Fig.~\ref{Distance}, the distance parameter of dynamic scatterers is smaller than that of static scatterers. This phenomenon is because the static scatterers, i.e., trees and buildings, are farther than dynamic scatterers, i.e., dynamic vehicles and UAVs surrounding the transceiver. 

\subsubsection{Angle Parameters}
\begin{figure*}[!t]
	\centering
	\subfigure[]{\includegraphics[width=0.32\textwidth]{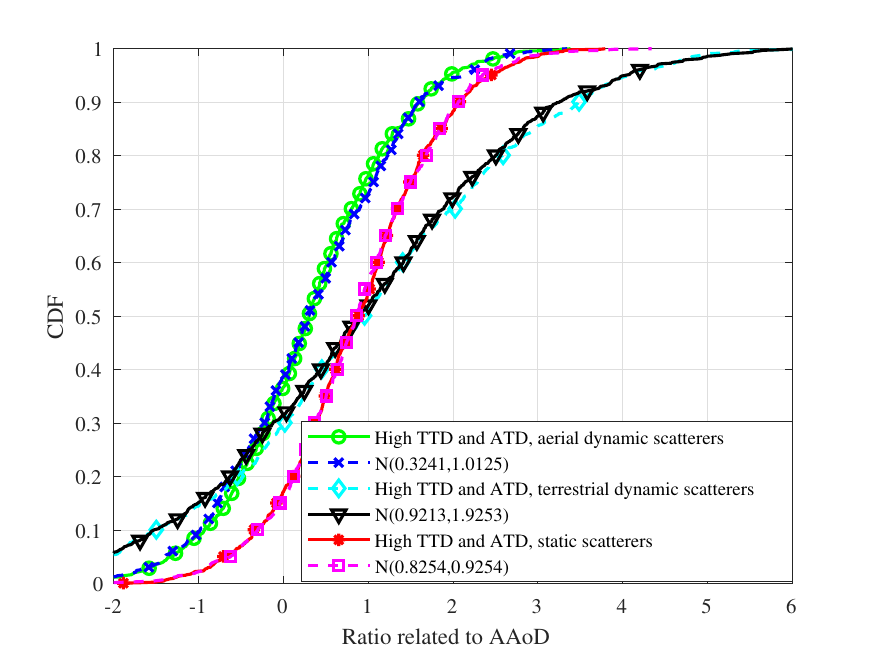}}
	\subfigure[]{\includegraphics[width=0.32\textwidth]{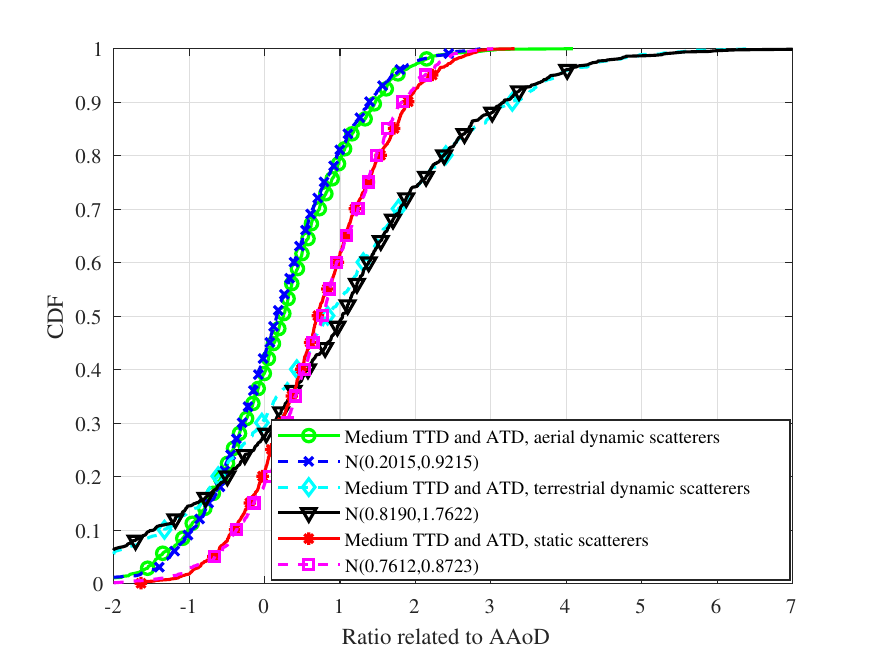}} 
   \subfigure[]{\includegraphics[width=0.32\textwidth]{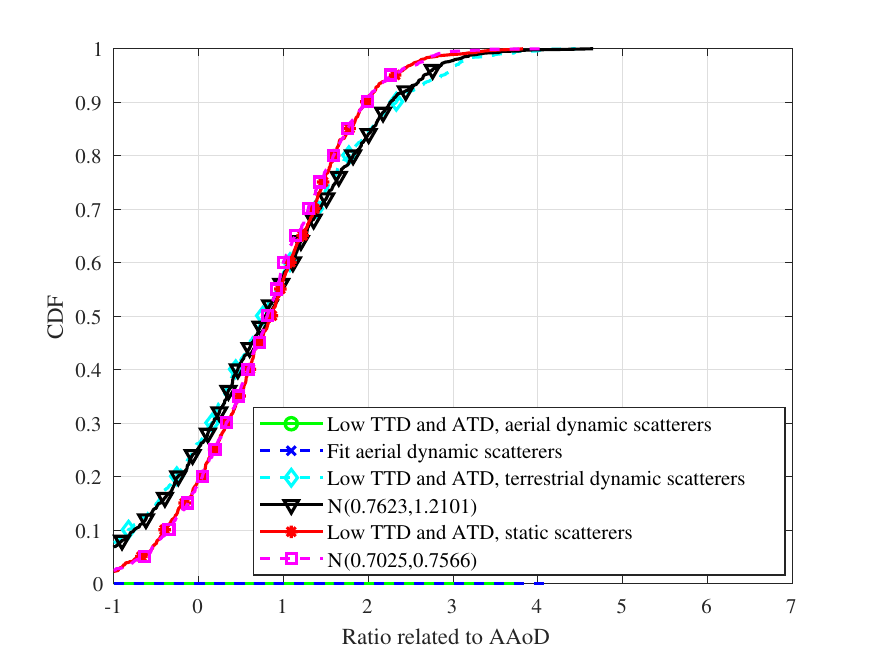}}
	\caption{CDFs of static/terrestrial dynamic/aerial dynamic scatterer angle parameters, i.e., AAoD, with the Gaussian distribution fitting under different TTD and ATD conditions.  Figs.~(a)--(c) show the CDFs of angle parameters under high, medium, and low TTD and ATD conditions, respectively.}
 \label{Angle}
\end{figure*}

There is currently no channel measurement or channel model considering the distinction among angle parameters of static, terrestrial dynamic, and aerial dynamic scatterers/clusters. The angle parameters in multi-UAV-to-multi-vehicle channels are for the first time analyzed under high, medium, and low TTD and ATD conditions, including azimuth angle of departure (AAoD), azimuth angle of arrival (AAoA), elevation angle of departure (EAoD), and elevation angle of arrival (EAoA) of static, terrestrial dynamic, aerial dynamic scatterers.
AAoD ratios of the $l/m/n$-th static/terrestrial dynamic/aerial dynamic scatterer in the transmission link from the $i$-th UAV to the $j$-th vehicle, 
${\alpha}^{\mathrm{U}_i,\mathrm{C}_j}_{\mathrm{S}_l}(t)$, ${\alpha}^{\mathrm{U}_i,\mathrm{C}_j}_{\mathrm{TD}_m}(t)$, and ${\alpha}^{\mathrm{U}_i,\mathrm{C}_j}_{\mathrm{AD}_n}(t)$, are expressed as
 \begin{equation}
      {\alpha}^{\mathrm{U}_i,\mathrm{C}_j}_{\mathrm{S}_l}(t)=\frac{\gamma^{\mathrm{U}_i,\mathrm{C}_j}_{\mathrm{S}_l}(t)}{\|\mathbf{T}^{\mathrm{U}_i}(t)-\mathbf{R}^{\mathrm{C}_j}(t)\|}
 \end{equation}
 \begin{equation}
     {\alpha}^{\mathrm{U}_i,\mathrm{C}_j}_{\mathrm{TD}_m}(t)=\frac{\gamma^{\mathrm{U}_i,\mathrm{C}_j}_{\mathrm{TD}_m}(t)}{\|\mathbf{T}^{\mathrm{U}_i}(t)-\mathbf{R}^{\mathrm{C}_j}(t)\|}
 \end{equation}
 \begin{equation}
     {\alpha}^{\mathrm{U}_i,\mathrm{C}_j}_{\mathrm{AD}_n}(t)=\frac{\gamma^{\mathrm{U}_i,\mathrm{C}_j}_{\mathrm{AD}_n}(t)}{\|\mathbf{T}^{\mathrm{U}_i}(t)-\mathbf{R}^{\mathrm{C}_j}(t)\|}
 \end{equation}
where $\gamma^{\mathrm{U}_i,\mathrm{C}_j}_{\mathrm{S}_l}(t)$/$\gamma^{\mathrm{U}_i,\mathrm{C}_j}_{\mathrm{TD}_m}(t)$/$\gamma^{\mathrm{U}_i,\mathrm{C}_j}_{\mathrm{AD}_n}(t)$ represents the AAoDs of the $l/m/n$-th static/terrestrial dynamic/aerial dynamic scatterer in the communication link from the $i$-th UAV to the $j$-th vehicle. Moreover, based on the constructed MUMV-CSCI dataset, the AAoDs of each static/terrestrial dynamic/aerial dynamic scatterer in each communication link at each snapshot are calculated and analyzed.
Figs.~\ref{Angle}(a)--(c) show the CDFs of all AAoDs of static, terrestrial dynamic, and aerial dynamic scatterers under high, medium, and low TTD and ATD conditions, respectively. The CDF of AAoDs matches the Gaussian distribution. The CDF of the Gaussian distribution for AAoDs related to static/terrestrial dynamic/aerial dynamic scatterers is represented by
\begin{equation}
    F^\mathrm{AAoD}_\mathrm{s/td/ad}(x)=\frac{1}{2}\left[1+\operatorname{erf}\left(\frac{x-\mu^\mathrm{AAoD}_\mathrm{s/td/ad}}{\sigma^\mathrm{AAoD}_\mathrm{s/td/ad} \sqrt{2}}\right)\right]
\end{equation}
where $\mu^\mathrm{AAoD}_\mathrm{s/td/ad}$ and $\sigma^\mathrm{AAoD}_\mathrm{s/td/ad}$ denote the mean value and the standard deviation of the 
Gaussian distribution for AAoDs related to static/terrestrial dynamic/aerial dynamic scatterers. $\operatorname{erf}(\cdot)$ is the error function. 
Similarly, the other static/terrestrial dynamic/aerial dynamic scatterer angle parameters, i.e., AAoA ${\theta}^{\mathrm{U}_i,\mathrm{C}_j}_{\mathrm{S/TD/AD}_{l/m/n}}(t)$, EAoD ${\beta}^{\mathrm{U}_i,\mathrm{C}_j}_{\mathrm{S/TD/AD}_{l/m/n}}(t)$, and EAoA ${\phi}^{\mathrm{U}_i,\mathrm{C}_j}_{\mathrm{S/TD/AD}_{l/m/n}}(t)$ are calculated in the same way, which also obey the Gaussian distribution and their corresponding statistical values are given in Table~\ref{Parameter_1}.
Compared with static scatterers, dynamic scatterers have larger variances in angle parameters. This is because that, the position of dynamic scatterers has more significant changes than that of static scatterers. Moreover, aerial dynamic scatterers have larger variances in angle parameters than terrestrial dynamic scatterers. This phenomenon is explained that UAVs have different flight heights, whereas vehicles on the ground are all located on the road at the same height. Furthermore, this phenomenon differs from the conclusions in vehicular communication presented in \cite{LAGBSM}, as the UAV's height has a significant impact on the distribution of scatterers.

\subsubsection{Power-Delay Characteristics}
\begin{figure*}[!t]
	\centering
	\subfigure[]{\includegraphics[width=0.32\textwidth]{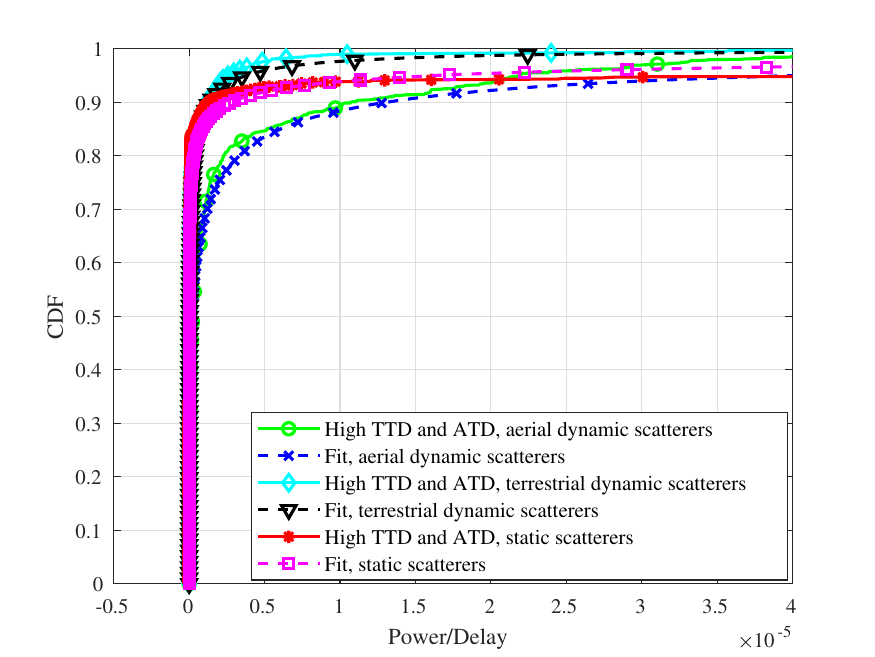}}
	\subfigure[]{\includegraphics[width=0.32\textwidth]{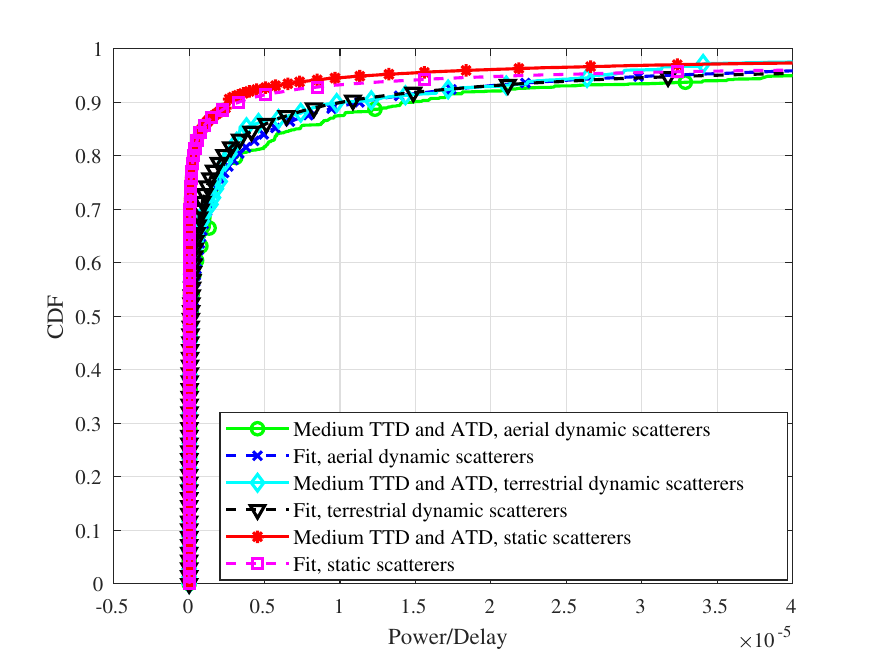}} 
   \subfigure[]{\includegraphics[width=0.32\textwidth]{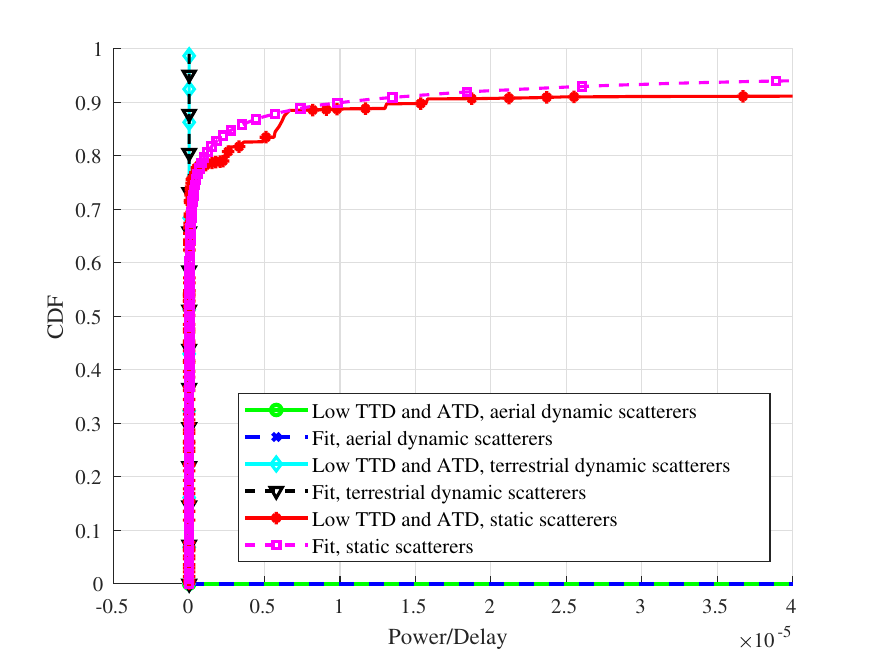}}
	\caption{CDFs of the ratios of static/terrestrial dynamic/aerial dynamic scatterer power to static/terrestrial dynamic/aerial dynamic scatterer delay with the Exponential expression fitting under different TTD and ATD conditions.  Figs.~(a)--(c) show the CDFs of power to delay parameters under high, medium, and low TTD and ATD conditions, respectively.}
 \label{Power}
\end{figure*}

The relationship between delay and power in multipath is a key channel characteristic in channel realization. In standardized model \cite{WINNER}, the path power is an exponential function of the path delay. The path power is separated into static, terrestrial dynamic, and aerial dynamic path power. The path power through $l/m/n$-th static/terrestrial dynamic/aerial dynamic scatterer is expressed by 
\begin{equation}
 P_{\mathrm{S}_{l}}(t)=\exp \left(-\xi_\mathrm{s} \tau_{\mathrm{S}_{l}}(t)-\eta_\mathrm{s}\right) 10^{-\frac{Z_\mathrm{s}}{10}} 
  \label{Po}
\end{equation}
\begin{equation}
 P_{\mathrm{TD}_{m}}(t)=\exp \left(-\xi_\mathrm{td} \tau_{\mathrm{TD}_{m}}(t)-\eta_\mathrm{td}\right) 10^{-\frac{Z_\mathrm{td}}{10}} 
  \label{Pp}
\end{equation}
\begin{equation}
 P_{\mathrm{AD}_{n}}(t)=\exp \left(-\xi_\mathrm{ad} \tau_{\mathrm{AD}_{n}}(t)-\eta_\mathrm{ad}\right) 10^{-\frac{Z_\mathrm{ad}}{10}} 
  \label{Pq}
\end{equation}
where $\xi_\mathrm{s/td/ad}$ and $\eta_\mathrm{s/td/ad}$ are the delay-related parameters of static/terrestrial dynamic/aerial dynamic scatterers. $\tau_{\mathrm{S/TD/AD}_{l/m/n}}(t)$ is the delay of the path through the $l/m/n$-th static/terrestrial dynamic/aerial dynamic scatterer. $Z_\mathrm{s/td/ad}$ follows the Gaussian distribution $\mathcal{N}\left(0, \sigma_\mathrm{E,s/td/ad}^2\right)$. 
For proper linear fitting, we transform \eqref{Po}, \eqref{Pp}, and \eqref{Pq} as
\begin{equation}
 -\mathrm{ln}P_{\mathrm{S}_{l}}(t)= \xi_\mathrm{s} \tau_{\mathrm{S}_{l}}(t)+\eta_\mathrm{s}+\frac{\ln 10}{10} Z_\mathrm{s} 
\end{equation}
\begin{equation}
 -\mathrm{ln}P_{\mathrm{TD}_{m}}(t)= \xi_\mathrm{td} \tau_{\mathrm{TD}_{m}}(t)+\eta_\mathrm{td}+\frac{\ln 10}{10} Z_\mathrm{td} 
\end{equation}
\begin{equation}
 -\mathrm{ln}P_{\mathrm{AD}_{n}}(t)= \xi_\mathrm{ad} \tau_{\mathrm{AD}_{n}}(t)+\eta_\mathrm{ad}+\frac{\ln 10}{10} Z_\mathrm{ad}. 
\end{equation}
The power and delay of each path through each static/terrestrial dynamic/aerial dynamic scatterer at each snapshot are calculated and fitted. The fitted parameters are summarized in Table~\ref{Parameter_1}. 
Figs.~\ref{Power}(a)--(c) show the fitting results under high, medium, and low TTD and ATD conditions,  which can validate the accuracy of the fitted parameters. Compared to static and terrestrial dynamic scatterers, the power of aerial dynamic scatterers is more sensitive to the change of delay, and thus the increase in the delay of aerial dynamic scatterers significantly reduces their power. 
\section{Multi-modal Intelligent Channel Model for 6G Multi-UAV-to-Multi-Vehicle Communications}
With the aid of LiDAR point clouds, static, terrestrial dynamic, and aerial dynamic scatterers can be distinguished. 
Based on the statistical distributions, a novel LiDAR-aided multi-UAV-to-multi-vehicle channel model is proposed, which considers the impact of different TTD and ATD conditions for the first time. Channel non-stationarity and consistency on the time and space domains and channel non-stationarity on the frequency domain are simultaneously depicted.
\subsection{Framework of the Proposed Multi-UAV-to-Multi-Vehicle Channel Model}
\begin{figure}[!t]
		\centering	\includegraphics[width=0.49\textwidth]{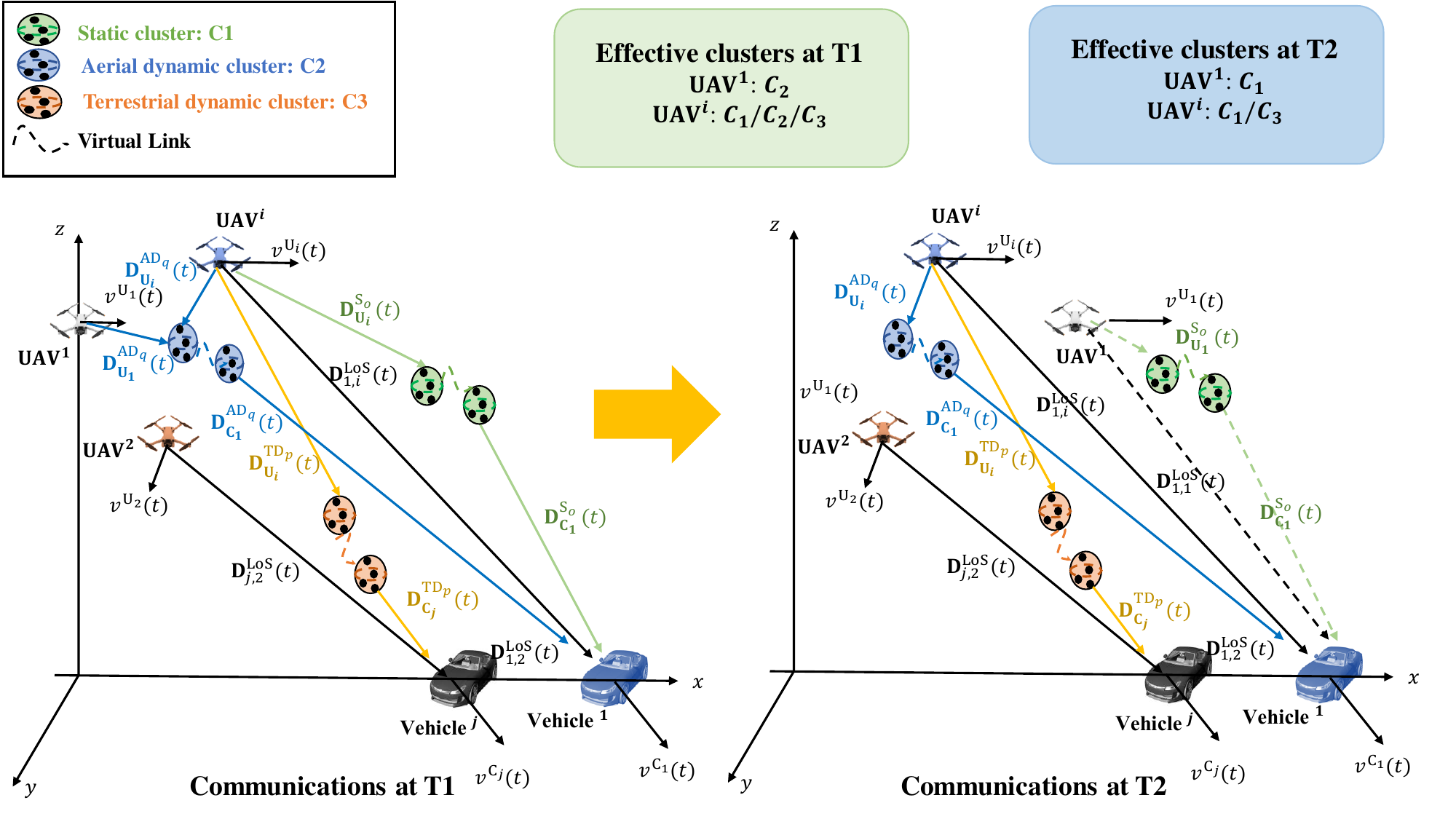}
	\caption{Geometry of the proposed channel model for multi-UAV-to-multi-vehicle intelligent sensing-communication integration and effective scatterers/clusters for the
transmission links at T1 and T2.}
	\label{fig:model}
	\end{figure}

In the proposed channel model, as shown in Fig.~\ref{fig:model}, the Txs and the Rxs are $I$ UAVs and $J$ vehicles, which are equipped with mmWave communication devices and LiDAR devices. 
The integrated channel impulse response (CIR) of the multi-UAV-to-multi-vehicle channel $H(t,\tau)$ is represented as
\begin{equation}
\begin{aligned} 
\mathbf{H}(t,\tau)=\left[
\begin{array}{cccc}
h_{1,1}(t,\tau) & h_{1,2}(t,\tau) &\cdots & h_{1,I}(t,\tau)\\
h_{2,1}(t,\tau) & h_{2,2}(t,\tau) &\cdots& h_{2,I}(t,\tau) \\
\vdots&\vdots&\ddots&\vdots\\
h_{J,1}(t,\tau) & h_{J,2}(t,\tau) &\cdots & h_{J,I}(t,\tau) \\
\end{array}
\right]
\end{aligned} 
\end{equation}
where the element $h_{j,i}(t,\tau)$, i.e., the CIR of transmission link from the $i$-th UAV to the $j$-th vehicle, is obtained by \eqref{desde}.
\begin{figure*}[!t]
	\begin{equation}
    \begin{aligned}
        h_{j,i}(t,\tau) &= \underbrace{\sqrt{\frac{\Omega_{ji}(t)}{\Omega_{ji}(t)+1}} h^\mathrm{LoS}_{j,i}(t)\delta\left(\tau-\tau^\mathrm{LoS}_{j,i}(t)\right)}_\mathrm{LoS} + \underbrace{\sqrt{\frac{\eta^\mathrm{GR}_{ji}(t)}{\Omega(t)+1}} h^\mathrm{GR}_{j,i}(t)\delta\left(\tau-\tau^\mathrm{GR}_{j,i}(t)\right)}_\mathrm{Ground \, Reflection} \\
        &+ \underbrace{\sum_{l=1}^{G^\mathrm{clu}(t)} \sum_{g_l=1}^{G^\mathrm{sca}(t)} \sqrt{\frac{\eta^\mathrm{NLoS}_{ji}(t)}{\Omega(t)+1}} h^{\mathrm{NLoS}_{l,g_l}}_{j,i}(t) \delta\left(\tau-\tau^{\mathrm{NLoS}_{l,g_l}}_{j,i}(t)\right)}_\mathrm{NLoS}.
    \end{aligned}
\label{desde}
\end{equation}
 		\hrulefill
\vspace*{4pt}
\end{figure*}
In \eqref{desde}, $\Omega_{ji}(t)$ represents Ricean factor of transmission link from the $i$-th UAV to the $j$-th vehicle. $\eta^\mathrm{GR}_{ji}(t)$ and $\eta^\mathrm{NLoS}_{ji}(t)$ are the power ratios of ground reflection component and NLoS component of transmission link from the $i$-th UAV to the $j$-th vehicle, as well as satisfy $\eta^\mathrm{GR}_{ji}(t)+\eta^\mathrm{NLoS}_{ji}(t)=1$. 
\subsubsection{For the LoS Component}
The LoS complex channel gain of transmission link from the $i$-th UAV to the $j$-th vehicle can be represented as
	\begin{equation}
	h^\mathrm{LoS}_{j,i}(t)=Q(t) \mathrm{exp}\left[j 2 \pi  \int_{t_0}^tf^\mathrm{LoS}_{j,i}(t)\mathrm{d}t+j \varphi^\mathrm{LoS}_{j,i}(t) \right]
	\label{LoS}
	\end{equation}
where $Q(t)$ is a rectangular window function \cite{window}. It is equal to 1 when $t_0\leqslant t \leqslant T_0$, otherwise it is equal to 0. The Doppler frequency, phase shift, and delay of LoS component of transmission link from the $i$-th UAV to the $j$-th vehicle are obtained by 
	\begin{equation}
    \begin{aligned}
    f^\mathrm{LoS}_{j,i}(t)=\frac{1}{\lambda} \frac{\left\langle\mathbf{D}^\mathrm{LoS}_{j,i}(t), \mathbf{v}^{\mathrm{C}_j}(t)-\mathbf{v}^{\mathrm{U}_i}(t)\right\rangle}{\left\|\mathbf{D}^\mathrm{LoS}_{j,i}(t)\right\|}
    \end{aligned}
 	\end{equation}
 	\begin{equation}
    \begin{aligned}
  \varphi^\mathrm{LoS}_{j,i}(t)=\varphi_{0}+\frac{2 \pi}{\lambda}\left\|\mathbf{D}^\mathrm{LoS}_{j,i}(t)\right\|
      \end{aligned}
  	\end{equation}
  	\begin{equation}
    \begin{aligned}
   \tau^\mathrm{LoS}_{j,i}(t)=\frac{\left\|\mathbf{D}^\mathrm{LoS}_{j,i}(t)\right\|}{c}
       \end{aligned}
   	\end{equation}
where $\left\langle \cdot,\cdot \right\rangle$, $\varphi_{0}$, and $\lambda$ are the inner product, initial phase shift, and carrier wavelength. $\mathbf{v}^{\mathrm{U}_i}(t)$ and $\mathbf{v}^{\mathrm{C}_j}(t)$ are the velocity vectors of the $i$-th UAV and the $j$-th vehicle. The distance vector from the $i$-th UAV to the $j$-th vehicle $\mathbf{D}^\mathrm{LoS}_{j,i}(t)$ is obtained by $\mathbf{D}^\mathrm{LoS}_{j,i}(t)=\mathbf{D}^\mathrm{LoS}_{j,i}(t_0)+\int^t_{t_0}\mathbf{v}^{\mathrm{C}_j}(t)\mathrm{d}t-\int^t_{t_0}\mathbf{v}^{\mathrm{U}_i}(t)\mathrm{d}t$.
\subsubsection{For the Ground Reflection Component}
The complex channel gain of transmission link from the $i$-th UAV to the $j$-th vehicle can be represented as
\begin{equation}
\begin{aligned}
&{h}^\mathrm{GR}_{j,i}(t)
=Q(t)\sqrt{P^\mathrm{GR}_{j,i}(t)}\\
&\times \mathrm{exp} \left\{j2\pi\left[\int_{t_0}^t{f}^\mathrm{GR,T}_{j,i}(t)\mathrm{d}t+\int_{t_0}^t{f}^\mathrm{GR,R}_{j,i}(t)\mathrm{d}t\right]+j \varphi^\mathrm{GR}_{j,i}(t)\right\}
\label{GRCIR}
\end{aligned}
\end{equation}
where $P^\mathrm{GR}_{j,i}(t)$, ${f}^\mathrm{GR,T/R}_{j,i}(t)$, $\varphi^\mathrm{GR}_{j,i}(t)$, and $\tau^\mathrm{GR}_{j,i}(t)$ denote power, Doppler frequency at the $i/j$-th UAV/vehicle, phase, and delay of ground reflection component from the $i$-th UAV to the $j$-th vehicle, respectively. Considering the limitation of paper length, these parameters can be calculated according to our previous work in \cite{bl1}.
\subsubsection{For the NLoS Component}The complex channel gain from the $i$-th UAV to the $j$-th vehicle via the $l$-th cluster by the $g_l$-th scatterer $h^{\mathrm{NLoS}_{l,g_l}}_{j,i}(t)$ is calculated by 

-- if the $l$-th cluster $\in$ $G^\mathrm{VR}_{j,i}(t)$ ($\forall o$)
	\begin{equation}
\begin{aligned}
    & h^{\mathrm{NLoS}_{l,g_l}}_{j,i}(t) = Q(t) \sqrt{P^{\mathrm{NLoS}_{l,g_l}}_{j,i}(t)} \\
    & \times \exp\left\{ j 2 \pi \left[ \int_{t_0}^t f^{\mathrm{T}_{l,g_l}}_{j,i}(t) \, \mathrm{d}t + \int_{t_0}^t f^{\mathrm{R}_{l,g_l}}_{j,i}(t) \, \mathrm{d}t \right] \right. \\
    & \quad + \left. j \varphi^{\mathrm{NLoS}_{l,g_l}}_{j,i}(t) \right\}
\end{aligned}
	\end{equation}

-- otherwise
\begin{gather}
h^{\mathrm{NLoS}_{l,g_l}}_{j,i}(t)=0
\end{gather}
where $G^\mathrm{VR}_{j,i}(t)$ is the set of visible twin-cluster in the transmission link from the $i$-th UAV to the $j$-th vehicle at time $t$, which can be obtained in Section III-B.
The Doppler frequency at Tx $f^{\mathrm{T}_{l,g_l}}_{j,i}(t)$, the Doppler frequency at Rx $f^{\mathrm{R}_{l,g_l}}_{j,i}(t)$, the phase shift $\varphi^{\mathrm{NLoS}_{l,g_l}}_{j,i}(t)$, the delay $\tau^{\mathrm{NLoS}_{l,g_l}}_{j,i}(t)$, and the distance $\mathbf{D}^{\mathrm{NLoS}_{l,g_l}}_{j,i}(t)$ are obtained by 
\begin{equation}
\begin{aligned}
 f^{\mathrm{T}_{l,g_l}}_{j,i}(t)=\frac{1}{\lambda} \frac{\left\langle\left(\mathbf{D}^{\mathrm{NLoS}_{l,g_l}}_{j,i}(t)\right), \mathbf{v}^{\mathrm{U}_i}(t)\right\rangle}{\left\|\mathbf{D}^{\mathrm{NLoS}_{l,g_l}}_{j,i}(t)\right\|}
\end{aligned}
\end{equation}
\begin{equation}
\begin{aligned}
 f^{\mathrm{R}_{l,g_l}}_{j,i}(t)=\frac{1}{\lambda} \frac{\left\langle\mathbf{D}(t)-\mathbf{D}^{\mathrm{NLoS}_{l,g_l}}_{j,i}(t), \mathbf{v}^{\mathrm{C}_j}(t)\right\rangle}{\left\|\mathbf{D}(t)-\mathbf{D}^{\mathrm{NLoS}_{l,g_l}}_{j,i}(t)\right\|}
\end{aligned}
\end{equation}
\begin{equation}
\begin{aligned}
&\varphi^{\mathrm{NLoS}_{l,g_l}}_{j,i}(t)=\varphi_{0}\\
&+\frac{2 \pi}{\lambda}\left[\left\|\mathbf{D}^{\mathrm{NLoS}_{l,g_l}}_{j,i}(t)\right\|+\left\|\mathbf{D}(t)-\mathbf{D}^{\mathrm{NLoS}_{l,g_l}}_{j,i}(t)\right\|+c \tilde{\tau}^{\mathrm{S}_{o}}(t)\right]
\end{aligned}
\end{equation}
\begin{equation}
\begin{aligned}
&\tau^{\mathrm{NLoS}_{l,g_l}}_{j,i}(t)\\
&=\frac{\left[\left\|\mathbf{D}^{\mathrm{NLoS}_{l,g_l}}_{j,i}(t)\right\|+\left\|\mathbf{D}(t)-\mathbf{D}^{\mathrm{NLoS}_{l,g_l}}_{j,i}(t)\right\|\right]}{c}+\tilde{\tau}^{\mathrm{S}_{o}}(t)
\end{aligned}
\end{equation}
\begin{equation}
\begin{aligned}
&\mathbf{D}^{\mathrm{NLoS}_{l,g_l}}_{j,i}(t)=\\
&{D}^{\mathrm{NLoS}_{l,g_l}}_{j,i}(t)\;
\left( \begin{array}{c}
\mathrm{cos}{\alpha}^{\mathrm{NLoS}_{l,g_l}}_{j,i}(t)\;\mathrm{cos}{\beta}^{\mathrm{NLoS}_{l,g_l}}_{j,i}(t) \\
\mathrm{sin}{\alpha}^{\mathrm{NLoS}_{l,g_l}}_{j,i}(t)\;\mathrm{cos}{\beta}^{\mathrm{NLoS}_{l,g_l}}_{j,i}(t) \\
\mathrm{sin}{\beta}^{\mathrm{NLoS}_{l,g_l}}_{j,i}(t)\\
\end{array}
 \right)
\end{aligned}
\end{equation}
where $\tilde{\tau}^{\mathrm{S}_{o}}(t)$ denotes the delay of virtual link in the $o$-th static cluster, which follows the Exponential distribution. The power parameter $P^{\mathrm{NLoS}_{l,g_l}}_{j,i}(t)$, and the distance parameter ${D}^{\mathrm{NLoS}_{l,g_l}}_{j,i}(t)$, the angle parameters ${\alpha}^{\mathrm{NLoS}_{l,g_l}}_{j,i}(t)$, and ${\beta}^{\mathrm{NLoS}_{l,g_l}}_{j,i}(t)$ are generated according to Table~\ref{Parameter_1}.

\subsection{Modeling of Channel Appearance and Disappearance in Multi-UAV-to-Multi-Vehicle Channel}
The objects, such as buildings, trees, UAVs, and vehicles, always exist. However, the scatterers/clusters are not effective for the transmission link if they are far away from transceivers. Based on the MUMV-CSCI dataset, as the analysis shown in Fig. \ref{fig:model}, it is obvious that the scatterers/clusters are not always effective. In the transmission links among different UAVs and vehicles, the sets of effective clusters are different, which results in the non-stationarity of scatterers/clusters on the space domain in the multi-UAV-to-multi-vehicle channel.
The appearance and disappearance of scatterers/clusters in the electromagnetic space are smooth as time and space evolve, resulting in the scatterer/cluster consistency on the time and space domains in the multi-UAV-to-multi-vehicle channel.
 
To accurately and simultaneously model the scatterer/cluster appearance and disappearance on the time and space domains in the multi-UAV-to-multi-vehicle channel, \textbf{a novel LiDAR-aided temporal and spatial non-stationarity and consistent algorithm} is developed as follows.

\textbf{Step 1: Initial setup of scatterers in the environment.} 
The parameters of static, terrestrial dynamic, and aerial dynamic scatterers under high, medium, and low TTD and ATD conditions are generated according to Table~\ref{Parameter_1}. For $\forall i,j$ ($i=1, 2, ..., I$; $j=1, 2, ..., J$), the numbers of static, terrestrial dynamic, and aerial dynamic scatterers between the $i$-th UAV and the $j$-th vehicle at initial time $t_0$ are generated by following the Logistic distribution. The distances at initial time $t_0$ are generated by following the Gamma distribution and Rayleigh distribution. The departure and arrival angles of each static, terrestrial dynamic, and aerial dynamic scatterers at initial time $t_0$, i.e., AAoDs, AAoAs, EAoDs, and EAoAs are generated by following the Gaussian distribution. According to the generated distances and angles of each static, terrestrial dynamic, and aerial dynamic scatterers, the locations of each static, terrestrial dynamic, and aerial dynamic scatterer at initial time $t_0$ are obtained. 

\textbf{Step 2: Obtaining all the clusters in the environment at initial time $t_0$.}
Based on the $K$-Means clustering algorithm, the generated static, terrestrial dynamic, and aerial dynamic scatterers are respectively clustered as static, terrestrial dynamic, and aerial dynamic clusters.

\textbf{Step 3: Obtaining the visible clusters for certain transmission links at initial time $t_0$.}
The visibility regions (VRs) of each UAV/vehicle are assumed as a semi-sphere with the center of the UAV/vehicle. The radii of the VR of the $i/j$-th UAV/vehicle is $R_\mathrm{vr}^{\mathrm{U}_i}$/$R_\mathrm{vr}^{\mathrm{C}_j}$, which is the maximum value of distances between initial generated static/terrestrial dynamic/aerial dynamic clusters and the $i/j$-th UAV/vehicle at initial time $t_0$, which is determined by Rayleigh distribution in Table~\ref{Parameter_1}. They are obtained as
	\begin{equation}
         \begin{aligned}
	R_\mathrm{vr}^{\mathrm{U}_i}=
 \mathop{\mathrm{max}}_{\forall o,q}\left\{ \left\|\mathbf{D}^{\mathrm{S}_{o}}_{\mathrm{U}_i}(t_0)\right\|,\left\|\mathbf{D}^{\mathrm{AD}_{q}}_{\mathrm{U}_i}(t_0)\right\|\right\}
	\label{R}
 \end{aligned}
	\end{equation}
 	\begin{equation}
 \begin{aligned}
	R_\mathrm{vr}^{\mathrm{C}_j}=\mathop{\mathrm{max}}_{\forall o,p}\left\{ \left\|\mathbf{D}^{\mathrm{S}_{o}}_{\mathrm{C}_j}(t_0)\right\|,\left\|\mathbf{D}^{\mathrm{TD}_{p}}_{\mathrm{C}_j}(t_0)\right\|\right\}
	\label{R}
 \end{aligned}
	\end{equation}
where $\mathbf{D}^{\mathrm{S}_{o}}_{\mathrm{U}_i/\mathrm{C}_j}(t_0)$/$\mathbf{D}^{\mathrm{TD}_{p}}_{\mathrm{U}_i/\mathrm{C}_j}(t_0)$/$\mathbf{D}^{\mathrm{TD}_{p}}_{\mathrm{U}_i/\mathrm{C}_j}(t_0)$ are the distance between the  $o/p/q$-th static/terrestrial dynamic/aerial dynamic cluster and the $i/j$-th UAV/vehicle at initial time $t_0$.
The cluster in the VRs of the $i/j$-th UAV/vehicle at time $t$, i.e., $R_\mathrm{vr}^{\mathrm{U}_i}$/$R_\mathrm{vr}^{\mathrm{C}_j}$ is the visible cluster for the $i/j$-th UAV/vehicle at time $t$. In this case, as the movement of UAVs and vehicles in the environment, the clusters are not always in the VRs of certain transceivers, which can represent the cluster appearance and disappearance on the time domain. As each transceiver, i.e., UAV/vehicle, has its own location and VR, different transmission links have different visible clusters, which represents the cluster appearance and disappearance on the space domain. Meanwhile, the VRs of UAVs and vehicles move as time evolves and share an integrated and consistent environment, leading to cluster consistency on the time and space domains.

\textbf{Step 4: Obtaining the visible clusters for certain transmission links at time $t=t_0+\Delta t$, including survived clusters and newly generated clusters.}
If the movement of clusters is still in the VRs of transceivers at time $t_0+\Delta t$, i.e., the distance between cluster and transceiver at time $t_0+\Delta t$ is still shorter than the radii of the VRs. The number of survived static/terrestrial dynamic/aerial dynamic clusters for the transmission between the $i$-th UAV and $j$-th vehicle at time $t_0+\Delta t$, i.e., visible static/terrestrial dynamic/aerial dynamic clusters at time $t_0$ that are still visible at time $t_0+\Delta t$, is denoted as $M^\mathrm{S}_\mathrm{s}(t_0+\Delta t)$/$M^\mathrm{S}_\mathrm{td}(t_0+\Delta t)$/$M^\mathrm{S}_\mathrm{ad}(t_0+\Delta t)$.
In addition to the survival clusters, there are some newly generated clusters for the transmission between the $i$-th UAV and the $j$-th vehicle at time $t_0+\Delta t$. 
For a certain distance between the $i$-th UAV and the $j$-th vehicle at time $t_0+\Delta t$, the number parameter  $M^\mathrm{L}_\mathrm{s}(t_0+\Delta t)$/$M^\mathrm{L}_\mathrm{td}(t_0+\Delta t)/M^\mathrm{L}_\mathrm{ad}(t_0+\Delta t)$ related to static/terrestrial dynamic/aerial dynamic clusters is randomly generated by obeying the Logistic distribution in Table~\ref{Parameter_1}. 
If the value $M^\mathrm{L}_\mathrm{s}(t_0+\Delta t)$/$M^\mathrm{L}_\mathrm{td}(t_0+\Delta t)/M^\mathrm{L}_\mathrm{ad}(t_0+\Delta t)$ is greater than the number of survived clusters at time $t_0+\Delta t$, i.e., $M^\mathrm{S}_\mathrm{s}(t_0+\Delta t)$/$M^\mathrm{S}_\mathrm{td}(t_0+\Delta t)$/$M^\mathrm{S}_\mathrm{ad}(t_0+\Delta t)$, the number of newly generated static/terrestrial dynamic/aerial dynamic clusters is given by
\begin{equation}
M^\mathrm{new}_\mathrm{s/td/ad}(t)=M^\mathrm{L}_\mathrm{s/td/ad}(t)-M^\mathrm{S}_\mathrm{s/td/ad}(t).    
\end{equation}
In this case, there are totally $M^\mathrm{L}_\mathrm{s}(t_0+\Delta t)$/$M^\mathrm{L}_\mathrm{td}(t_0+\Delta t)/M^\mathrm{L}_\mathrm{ad}(t_0+\Delta t)$ static/terrestrial dynamic/aerial dynamic clusters that contribute to channel realization.
On the contrary, if $M^\mathrm{L}_\mathrm{s}(t_0+\Delta t)$/$M^\mathrm{L}_\mathrm{td}(t_0+\Delta t)/M^\mathrm{L}_\mathrm{ad}(t_0+\Delta t)$ is less than $M^\mathrm{S}_\mathrm{s}(t_0+\Delta t)$/$M^\mathrm{S}_\mathrm{td}(t_0+\Delta t)$/$M^\mathrm{S}_\mathrm{ad}(t_0+\Delta t)$,
the number of newly generated clusters is equal to zero, i.e., $M^\mathrm{new}_\mathrm{s/td/ad}(t)=0$. In this case, there are $M^\mathrm{S}_\mathrm{s}(t_0+\Delta t)$/$M^\mathrm{S}_\mathrm{td}(t_0+\Delta t)$/$M^\mathrm{S}_\mathrm{ad}(t_0+\Delta t)$ static/terrestrial dynamic/aerial clusters that contribute to channel realization. 

\textbf{Step 5: Randomly matching the mixed twin-clusters.}
The visible static and aerial dynamic clusters for the transmission link from the $i$-th UAV to the $j$-th vehicle at time $t_0+\Delta t$ are the sub-cluster around Tx, i.e., the $i$-th UAV, and the visible static and terrestrial dynamic clusters for the transmission link from the $i$-th UAV to the $j$-th vehicle at time $t_0+\Delta t$ are the sub-cluster around Rx, i.e., the $j$-th vehicle. The sub-cluster around the $i$-th UAV to the $j$-th vehicle is matched randomly as the set of visible twin-cluster in the transmission link from the $i$-th UAV to the $j$-th vehicle at time $t_0+\Delta t$, i.e., $G^\mathrm{VR}_{j.i}(t_0+\Delta t)$.

\textbf{Step 6: Modeling the channel non-stationarity on the frequency domain.} 
The CIR of transmission link from the $i$-th UAV to the $j$-th vehicle at time $t=t_0+\Delta t$, i.e., $h_{j,i}(t, \tau)$, is obtained by \eqref{desde}. The Fourier transform of $h_{j,i}(t, \tau)$ in respect of $\tau$, i.e., the time-varying transfer function $H_{j,i}'(t, f)$ is calculated, which is expressed as
\begin{equation}
 H_{j,i}'(t, f)=\int_{-\infty}^{\infty} h(t, \tau) \mathrm{exp}\left({-j 2 \pi f \tau}\right) \mathrm{d} \tau.   
\end{equation}
Considering the frequency-dependent factor $\left(\frac{f}{f_{c}}\right)^{\chi}$, the time-varying transfer function is calculated by \eqref{eee},
\begin{figure*}[!t]
	\begin{equation}
 	\begin{aligned}
	H_{j,i}(t, f)= &\underbrace{\sqrt{\frac{\Omega(t)}{\Omega(t)+1}} h^{\mathrm{LoS}}_{j,i}(t) \mathrm{exp}\left[{-j 2 \pi f \tau^{\mathrm{LoS}}_{j,i}(t)}\right]}_\mathrm{LoS} +\underbrace{\sqrt{\frac{\eta^\mathrm{GR}(t)}{\Omega(t)+1}} \left(\frac{f}{f_{c}}\right)^{\chi} h^\mathrm{GR}_{j,i}(t) \mathrm{exp}\left[{-j 2 \pi f\tau^\mathrm{GR}_{j,i}(t)}\right]}_\mathrm{Ground \, Reflection}\\
&+\underbrace{\sqrt{\frac{\eta^\mathrm{NLoS}(t)}{\Omega(t)+1}}\left(\frac{f}{f_{c}}\right)^{\chi}\sum_{l=1}^{G^\mathrm{clu}(t)}\sum_{g_l=1}^{G^\mathrm{sca}}h^{\mathrm{NLoS}_{l,g_l}}_{j,i}(t)\mathrm{exp}\left[{-j 2 \pi f\tau^{\mathrm{NLoS}_{l,g_l}}_{j,i}}\right]	}_\mathrm{NLoS}
 \label{eee}
		\end{aligned}
	\end{equation}
 		\hrulefill
\vspace*{4pt}
\end{figure*}
where $\chi$ is the frequency-dependent parameter \cite{ewf2w}.

\textbf{Cycling Step 4- Step 6 by $t=t+\Delta t$.}

\section{Channel Statistical Properties}
In this section, the key statistical properties for the proposed multi-UAV-to-multi-vehicle channel model are obtained, including the TSF-CF, TSI, and DPSD.

\subsection{Space-Time-Frequency Correlation Function}
The TSF-CF of the transmission from the $i$-th UAV to the $j$-th vehicle on the ground can be calculated as \cite{property} 

\begin{equation}
R_{ji,j'i'}(t,f;\Delta{t},\Delta{f})=\mathbb{E}[h_{ji}^\ast(t,f)h_{j'i'}(t+\Delta{t},f+\Delta{f})]
\label{eq:CF}
\end{equation}where $\mathbb{E}[\cdot]$ and $(\cdot)^*$ denote the expectation operation and complex conjugate operation, respectively. Since the TSF-CFs of LoS component, ground reflection component, and NLoS component can be assumed as independent of each other, the TSF-CF can be further obtained by the sum of the TSF-CFs of LoS component, ground reflection component, and NLoS component, i.e., (\ref{eq:CF_sum}),
\begin{figure*}
\begin{equation}
	\begin{aligned}
R_{ji,j'i'}(t,f;\Delta{t},\Delta{f})&=R_{ji, j'i'}^\mathrm{LoS}(t,f;\Delta{t},\Delta{f})+R_{ji, j'i'}^\mathrm{GR}(t,f;\Delta{t},\Delta{f})+R_{ji, j'i'}^\mathrm{NLoS}(t,f;\Delta{t},\Delta{f})
	\label{eq:CF_sum}
		\end{aligned}
	\end{equation}
\end{figure*}where the correlation of LoS component, ground reflection component, and NLoS component can be computed as (\ref{eq:zLoS})--(\ref{eq:z1NLoS}).
\begin{figure*}
\begin{equation}
\label{eq:zLoS}
\begin{aligned}
	R_{ji, j'i'}^\mathrm{LoS}(t,f;\Delta{t},\Delta{f})=\sqrt{\frac{\Omega_{ji}(t)}{\Omega_{ji}(t)+1}\frac{\Omega_{j'i'}(t+\Delta{t})}{\Omega_{j'i'}(t+\Delta{t})+1}}
	{h_{j,i}^{\mathrm{LoS\ast}}(t)h_{j',i'}^{\mathrm{LoS}}(t+\Delta{t})e^{j2\pi \left( f\tau_{j,i}^{\mathrm{LoS}}(t)-(f+\Delta{f})\tau_{j',i'}^{
 \mathrm{LoS}}(t+\Delta{t})\right)}}
	\end{aligned}
	\end{equation}
 
\begin{equation}
\begin{aligned}
R_{ji, j'i'}^\mathrm{GR}(t,f;\Delta{t},\Delta{f})=\sqrt{\frac{\eta^\mathrm{GR}_{ji}(t)}{\Omega_{ji}(t)+1}\frac{\eta^\mathrm{GR}_{j',i'}(t+\Delta{t})}{\Omega_{j'i'}(t+\Delta{t})+1}}
h_{j,i}^{\mathrm{GR\ast}}(t)h_{j',i'}^{\mathrm{GR}}(t+\Delta{t})e^{j2\pi \left(f\tau_{j,i}^{\mathrm{GR}}(t)-(f+\Delta{f})\tau_{j',i'}^{\mathrm{GR}}(t+\Delta{t})\right)}
\end{aligned}
\label{eq:zNLoS}
\end{equation}
 
\begin{equation}
\begin{aligned}
    R_{ji, j'i'}^\mathrm{NLoS}(t,f;\Delta{t},\Delta{f}) &= \sqrt{\frac{\eta^\mathrm{NLoS}_{ji}(t)}{\Omega_{ji}(t)+1} \cdot \frac{\eta^\mathrm{NLoS}_{j'i'}(t+\Delta{t})}{\Omega_{j'i'}(t+\Delta{t})+1}} \times \mathbb{E}\left[ \sum_{l=1}^{G^\mathrm{clu}(t)} \sum_{l'=1}^{G^\mathrm{clu}(t+\Delta{t})} \sum_{g_l=1}^{G^\mathrm{sca}(t)} \sum_{g'_l=1}^{G^\mathrm{sca}(t+\Delta{t})} \right. \\
    & \quad \left. h_{j,i}^{\mathrm{NLoS}_{l,g_l\ast}}(t) h_{j',i'}^{\mathrm{NLoS}_{l',g'_l}}(t+\Delta{t}) e^{j2\pi \left( f \tau_{j,i}^{\mathrm{NLoS}_{l,g_l}}(t) - (f+\Delta{f}) \tau_{j',i'}^{\mathrm{NLoS}_{l',g'_l}}(t+\Delta{t}) \right)} \right].
\end{aligned}
\label{eq:z1NLoS}
\end{equation}

   		\hrulefill
\vspace*{4pt}
\end{figure*}
For a certain UAV, the TSF-CFs can be simplified to the cooperative space cross-correlation function (CCF) between different vehicles by setting $i=i'$, $j \neq j'$, $\Delta t = 0$, and $\Delta f = 0$.
For a certain vehicle, the TSF-CFs can be simplified to the space CCF between different UAVs by setting $j = j'$, $i \neq i'$, $\Delta t = 0$, and $\Delta f = 0$.
The TSF-CF can be simplified to the time auto-correlation function (ACF) by setting $i = i'$, $j = j'$, and $\Delta f = 0$.
Furthermore, the TSF-CF can be simplified to the frequency correlation function (FCF) by setting $i = i'$, $j = j'$, and $\Delta t = 0$.

\subsection{Time Stationary Interval}
If the absolute value of the relative error of the delay spread is not more than 10\%, the CIR can be regarded as stationary \cite{0808}. In this case, the corresponding minimum time interval of stationary CIR is TSI.
The TSI of the proposed multi-UAV-to-multi-vehicle channel model is obtained by
 
\begin{equation}
T_s(t)=\mathrm{inf}\{\Delta t|_{\frac{\left\|A^\mathrm{(2)}_{\tau'}(t+\Delta t)-A^\mathrm{(2)}_{\tau'}(t)\right\|}{A^\mathrm{(2)}_{\tau'}(t)}\leq 0.1}\}
\end{equation}
 where inf$\{\cdot\}$ is the infimum of a certain function. $A^\mathrm{(2)}_{\tau'}(t)$ denotes the time-variant delay spread and can be obtained by (\ref{timeinter}).
\begin{figure*}
\begin{equation}
\begin{aligned}
&A^\mathrm{(2)}_{\tau'}(t) = \\
&\sqrt{
\frac{
\sum^{J}_{j=1} \sum^{I}_{i=1} \sum_{l=1}^{G^\mathrm{clu}(t)} \sum_{g_l=1}^{G^\mathrm{sca}(t)} (c_{ji,l,g_l}(t))^2 (\tau_{j,i}^{\mathrm{NLoS}_{l,g_l}}(t))^2
}{
\sum^{J}_{j=1} \sum^{I}_{i=1} \sum_{l=1}^{G^\mathrm{clu}(t)} \sum_{g_l=1}^{G^\mathrm{sca}(t)} (c_{ji,l,g_l}(t))^2
} 
- \left( 
\frac{
\sum^{J}_{j=1} \sum^{I}_{i=1} \sum_{l=1}^{G^\mathrm{clu}(t)} \sum_{g_l=1}^{G^\mathrm{sca}(t)} (c_{ji,l,g_l}(t))^2 \tau_{j,i}^{\mathrm{NLoS}_{l,g_l}}(t)
}{
\sum^{J}_{j=1} \sum^{I}_{i=1} \sum_{l=1}^{G^\mathrm{clu}(t)} \sum_{g_l=1}^{G^\mathrm{sca}(t)} (c_{ji,l,g_l}(t))^2
} 
\right)^2.
}
\end{aligned}
\label{timeinter}
\end{equation}

 					\hrulefill
\vspace*{4pt}
\end{figure*}
In \ref{timeinter}, $c_{ji,l,g_l}$ is the path gain of the $g_l$-th ray in the $l$-th twin-cluster between the $i$-th UAV and the $j$-th vehicle.

\subsection{Doppler Power Spectral Density}
Based on the Fourier transform of the TACF, the DPSD can be obtained by
	\begin{equation}
            \begin{aligned}
	\Upsilon(t;f_\mathrm{D})=\int_{-\infty}^{+\infty}\zeta(t;\Delta{t})e^{-j2\pi{f_\mathrm{D}}\Delta{t}}\mathrm{d}(\Delta{t})
	\end{aligned}
	\end{equation}
where $f_\mathrm{D}$ and $\zeta(t;\Delta{t})$ denote the Doppler frequency and TACF. The time-varying DPSD illustrates the time-varying characteristic of the proposed channel.
\begin{figure}[!t]
    \centering
    \includegraphics[width=0.49\textwidth]{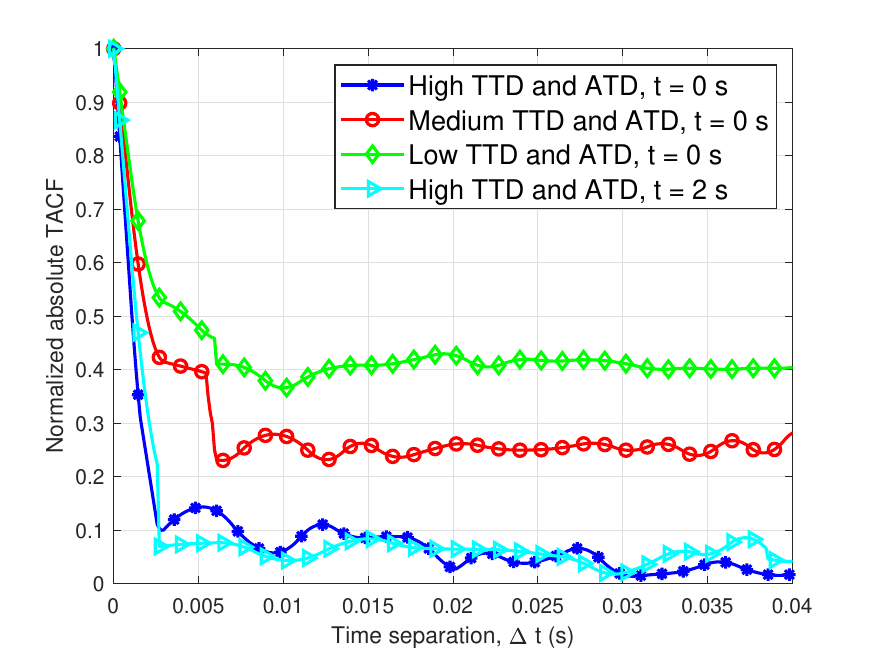}
    \caption{TACFs under different TTD and ATD conditions and different time instants.}
    \label{TACF}
\end{figure}

\begin{figure}[!t]
    \centering
    \includegraphics[width=0.49\textwidth]{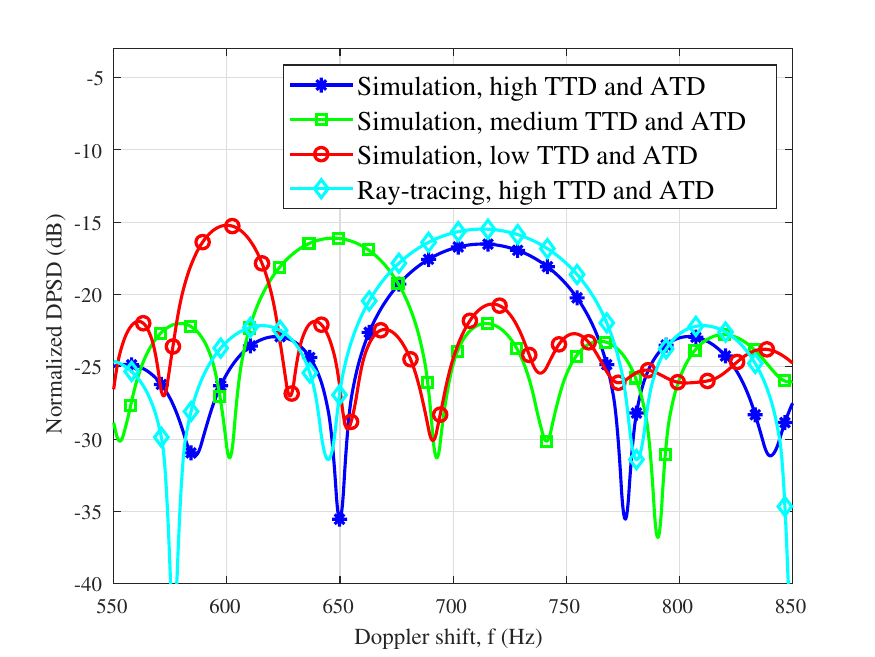}
    \caption{Comparison of simulated DPSDs and RT-based DPSDs under different TTD and ATD conditions.}
    \label{DPSD}
\end{figure}

\begin{figure}[!t]
    \centering
    \includegraphics[width=0.49\textwidth]{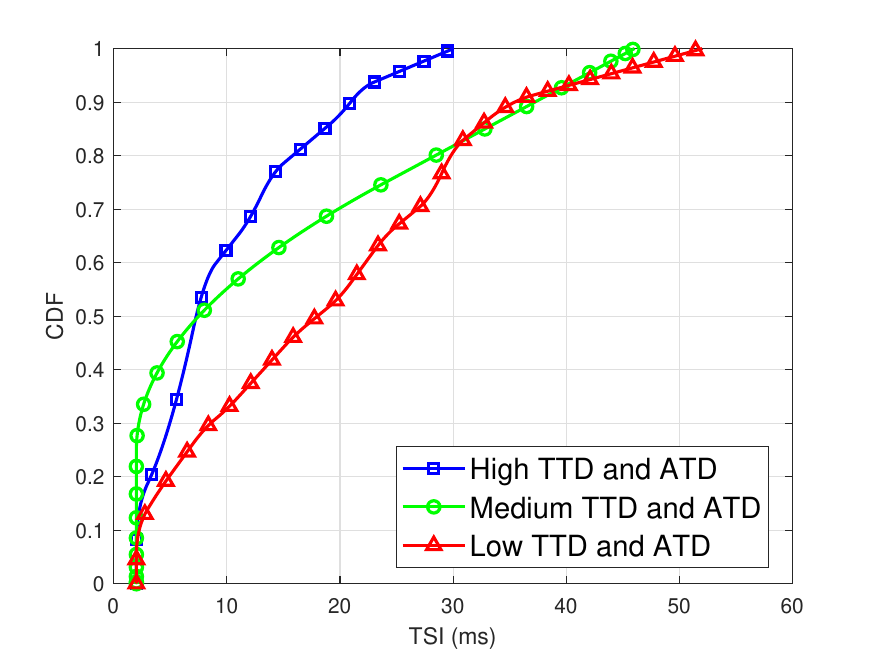}
    \caption{CDFs of TSIs under different TTD and ATD conditions.}
    \label{TSI}
\end{figure}

\section{Simulation Results and Analysis}
The key statistical properties of the channels are simulated and compared with precise RT-based results. Carrier frequency is $f_\mathrm{c}=28$~GHz and the bandwidth is $2$~GHz. 
The azimuth and elevation angles of the Tx and Rx are $\phi_\mathrm{T}^\mathrm{E}=\phi_\mathrm{R}^\mathrm{E}=\pi/4$,  $\theta_\mathrm{T}^\mathrm{A}=\pi/3$, and $\theta_\mathrm{R}^\mathrm{A}=3\pi/4$. Delays of virtual links $\tau_{i}(t)$ and $\tau_{j}(t)$ obey the Exponential distribution with the mean and variance 80 ns and 15 ns to imitate the complex transmission between twin-clusters. The environment-dependent factor is set to $\chi=1.35$ \cite{ewf2w}. The aforementioned parameters remain unchanged unless otherwise stated.

Fig. \ref{TACF} shows the absolute normalized TACFs under low, medium, and high TTD and ATD conditions, at $t = 0$~s and $t = 2$~s. From Fig. \ref{TACF}, TACFs depend on time instants and time separations. Meanwhile, time non-stationarity is depicted. In addition, the TACF decreases as the TTD and ATD increase. This is because that, as the number of vehicles and UAVs increases, the channel becomes more variable and the temporal correlation decreases.

We obtain RT-based CIRs collected in Wireless InSite with the scenario shown in Fig. \ref{WI_AirSim}. As shown in Fig. \ref{DPSD}, DPSD is derived based on the CIR data under the high TTD and ATD conditions and is further compared with the simulated DPSD in high, medium, and low TTD and ATD conditions. In Fig. \ref{DPSD}, in high TTD and ATD conditions, the RT-based DPSD is much closer to the simulated DPSD. The DPSD is flatter in high and medium TTD and ATD conditions compared to low TTD and ATD conditions. Since UAVs and vehicles are denser and channels are more complex in high TTD and ATD conditions. Therefore, the comparison of different TTD and ATD conditions is significant for the proposed multi-UAV-to-multi-vehicle channel model.

Fig. \ref{TSI} presents the CDFs of channel TSIs under different TTD and ATD conditions. In Fig. \ref{TSI}, the TSI of the multi-UAV-to-multi-vehicle channel decreases as TTD and ATD conditions increase. This is attributable to the fact that more UAVs and vehicles lead to a more complex multi-UAV-to-multi-vehicle channel. Therefore, the multi-UAV-to-multi-vehicle channel under high TTD and ATD conditions is more sophisticated and more variable, which results in a lower TSI.

\section{Conclusions}
This paper has proposed a novel multi-modal intelligent channel model for 6G multi-UAV-to-multi-vehicle communications. The proposed model has incorporated both TTD and ATD, which can capture channel non-stationarity and the consistent nature of the channel on time, space, and frequency domains. A new MUMV-CSCI dataset, including channel information and LiDAR point clouds, has been constructed to parameterize the proposed model under different TTD and ATD conditions. The proposed model has accurately characterized static scatterers, terrestrial dynamic scatterers, and aerial dynamic
scatterer and utilized statistical distributions to describe the properties. Simulation results, validated against the RT-based data, have demonstrated the model's ability to capture key channel statistics and its suitability for future 6G low-altitude transportation communication systems.

\ifCLASSOPTIONcaptionsoff
  \newpage
\fi

\end{document}